\documentclass[prb,aps,twocolumn,showpacs]{revtex4}

\usepackage{graphicx,epsfig}% Include figure files
\usepackage{bm}% bold math
\usepackage{amsmath}

\begin{document}

\title{Single-molecule-mediated heat current between
       an electronic and a bosonic bath}
\smallskip

\author{Yuval Vinkler-Aviv,$^1$ Avraham Schiller,$^{1,} \footnote{Deceased, June 22nd 2013.}$
        and Natan Andrei$^2$}

\affiliation{$^1$Racah Institute of Physics,
                 The Hebrew University, Jerusalem 91904,
                 Israel\\
             $^2$Center for Materials Theory, Department
                 of Physics, Rutgers University,
                 Piscataway, NJ 08854-8019 USA}

\begin{abstract}
In molecular devices electronic degrees of freedom are coupled
to vibrational modes of the molecule, offering
an opportunity to study fundamental
aspects of this coupling at the nanoscale. To this end we 
consider the nonequilibrium heat
exchange between a conduction band and a bosonic
bath mediated by a single molecule. For molecules large enough so that on-site Coulomb repulsion can be dropped, we carry out an asymptotically exact calculation of the heat current, governed by the smallness of the electron-phonon coupling, and obtain the steady state heat current  driven by a finite temperature drop. At low temperatures the heat current   is found to have a power-law behavior  with respect to the temperature difference  with the power depending on the nature of the bosonic bath. At high temperatures, on the other hand, the current is linear in the temperature difference for all types of bosonic baths. The crossover between these behaviors is described. Some of the results are given a physical explanation by comparing to a perturbative Master equation calculation (whose limitation we examine).
\end{abstract}


\maketitle

\section{Introduction}

Recent developments in the fabrication and control of nanostructures and molecular devices have stimulated a growing interest in the study and research of heat conductance in such devices.~\cite{Reviews-of-SMTs,Wang2008,Phillpot2003,Natelson2011} As efforts are being made to better utilize and control these devices, the understandings of the mechanisms for accumulation and dissipation of heat are of importance. The literature on the topic considers different physical setups, differing in the process by which heat is mediated, the devices under investigation and the nature of the heat baths involved in the heat transfer process.~\cite{Galperin2007,Entin2010,Entin2012,Ozpineci2001,Segal2003,Segal2011,Buttiker2012}

We consider heat transfer  between two baths held at different temperatures  through a steady-state current of energy. The linear response regime, which relies on the equilibrium properties of the system, arises in the limit of small gradient of  temperatures. A linear response study of a system similar to the one that is presented in this paper was done recently by Entin-Wohlman and others and the thermopower properties were calculated~\cite{Entin2010,Entin2012}. In the general case, however, the system is far from equilibrium, as the temperature gradient is finite. The description of such systems is a challenging problem of great interest in current research, as many of the concepts and techniques used to describe equilibrium setups are inadequate. Exact solutions of systems far from equilibrium are particularly desirable as they may offer both a benchmark and an unbiased understanding of the underlying physics.

In this paper we present an asymptotically-exact calculation of the heat current through a molecular junction, under explicit nonequilibrium conditions manifested by a finite temperature gradient between the two baths to which the molecule is coupled -- the electrons in a conduction band and a bosonic bath.  In a typical molecular bridge, molecular orbitals are coupled simultaneously to the lead electrons and to the vibrational modes of the molecule, with the former degrees of freedom reduced to a single effective band in the absence of a bias voltage.~\cite{Glazman-Raikh-88} This coupling to the vibrational modes of the molecule is believed to have an essential role in heat transfer processes.~\cite{Natelson2011,Galperin2007,Entin2010,Entin2012,Franke2012,Kawai2010}
In the continuum limit, a minimal model for an unbiased molecular bridge therefore consists of a vibrational mode that is coupled by displacement to the conduction electrons at the origin, and is also coupled to a bosonic bath of vibrational modes, as described by the Hamiltonian of Eq.~(\ref{H}). The solution is asymptotically exact in the sense that it is governed by the smallness of the electron-phonon coupling $g$ with respect to the effective energy band of the conductance electrons.

The thermal properties of the bosonic Hamiltonian of eq.~(\ref{H-bosonic}), and similar Hamiltonians describing heat current between harmonic baths, were studied before in different physical setups than presented here~\cite{Segal2008,Ojanen2008,Ojanen2011,Dahr2008,Saito2013}. The results we derive in this paper corroborate those studies, and adjust them to describe the electronic systems which we shall present.

The paper is organized as follows. We begin in Sec.~\ref{sec:model} by presenting the model and the physical systems which it may represent, under suitable mappings. In Sec.~\ref{sec:mapping} we then introduce the nonequilibrium condition and map the model onto a form quadratic in bosons. This quadratic nature of the Hamiltonian is then exploited, in Sec.~\ref{sec:resonance_heat_current}, in order to calculate exactly the heat current in the system. In Sec.~\ref{sec:master_equation} we then turn to a Master equation approach, which is perturbative in nature, in order to calculate again the heat current. This approach allows us to gain some useful physical understanding of the processes involved. Section~\ref{sec:off-resonance} addresses the case where the localized fermionic level is taken off-resonance, which breaks the particle-hole symmetry of the problem and adds a linear term to our quadratic Hamiltonian. We then turn, in Sec.~\ref{sec:conclusions} to present our conclusions.

\section{The Model}
\label{sec:model}

\begin{figure}[tbp]
\begin{center}
\includegraphics[width=0.44\textwidth]{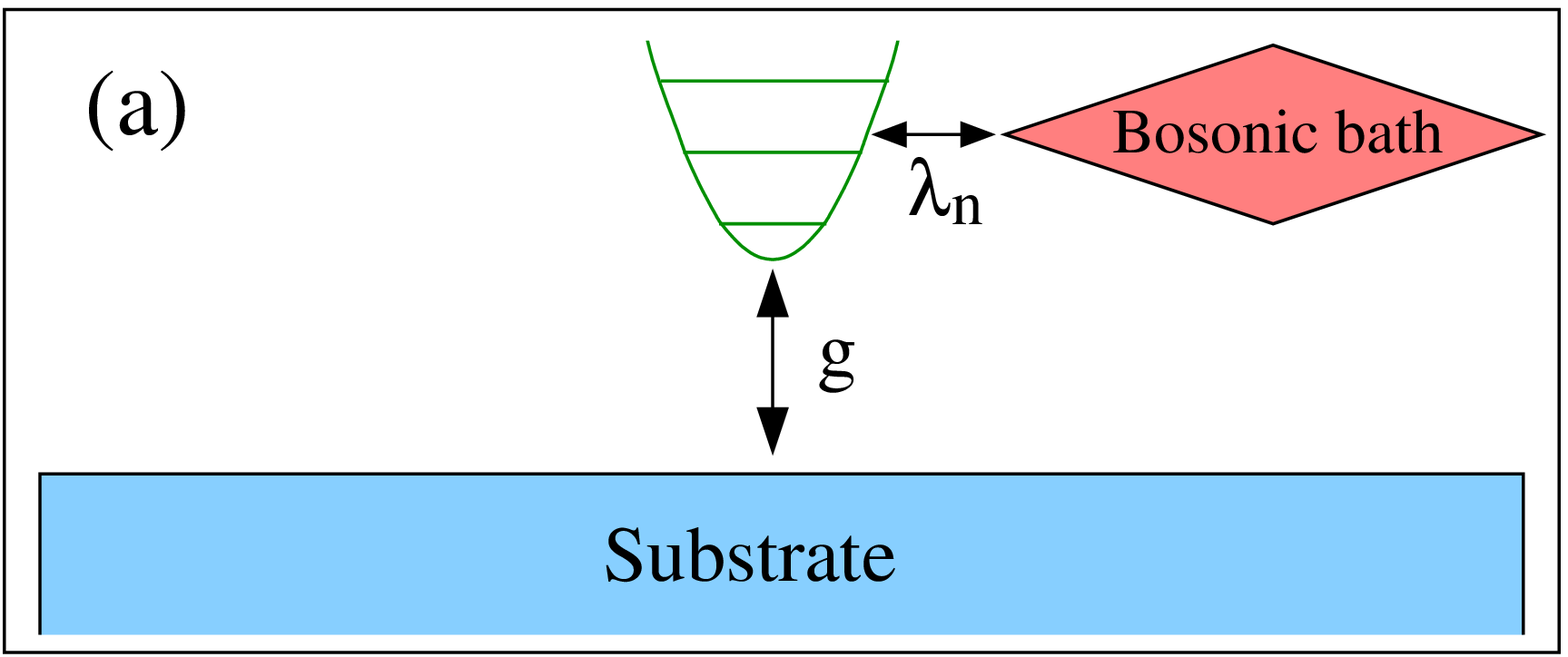}
\includegraphics[width=0.44\textwidth]{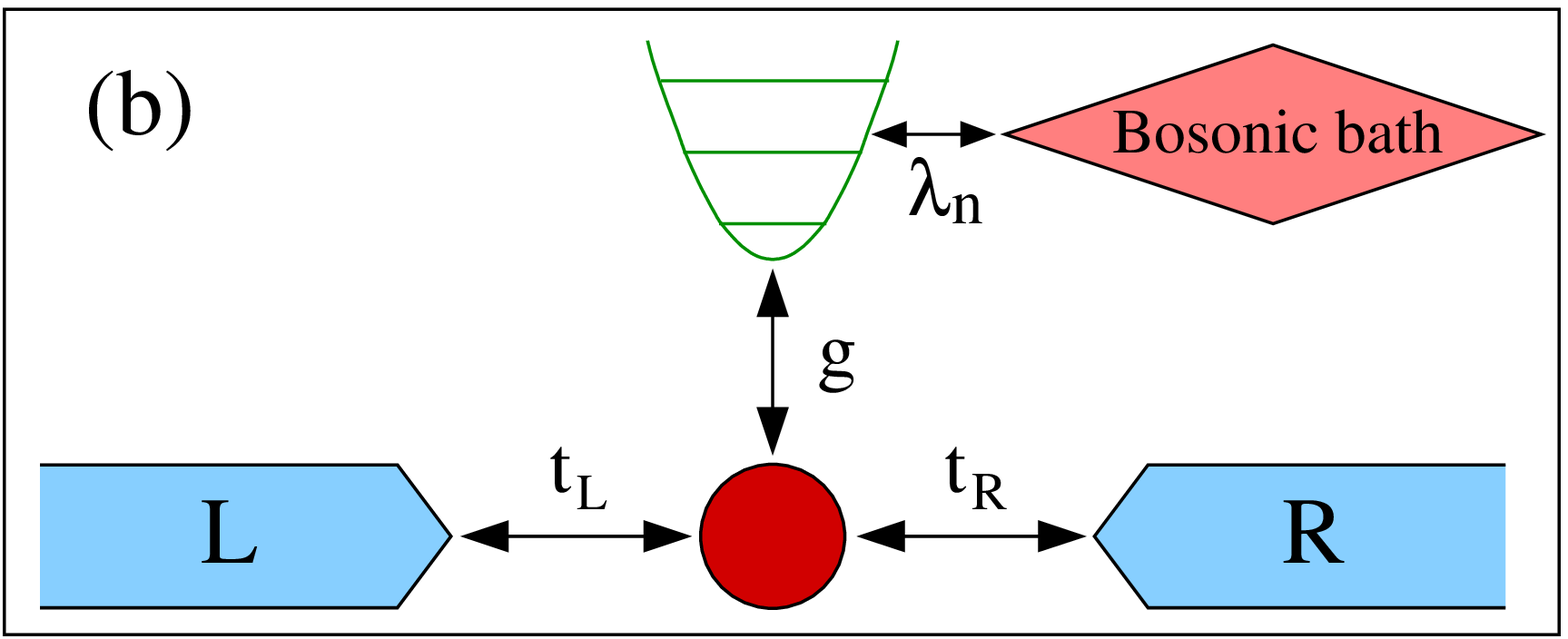}
\includegraphics[width=0.44\textwidth]{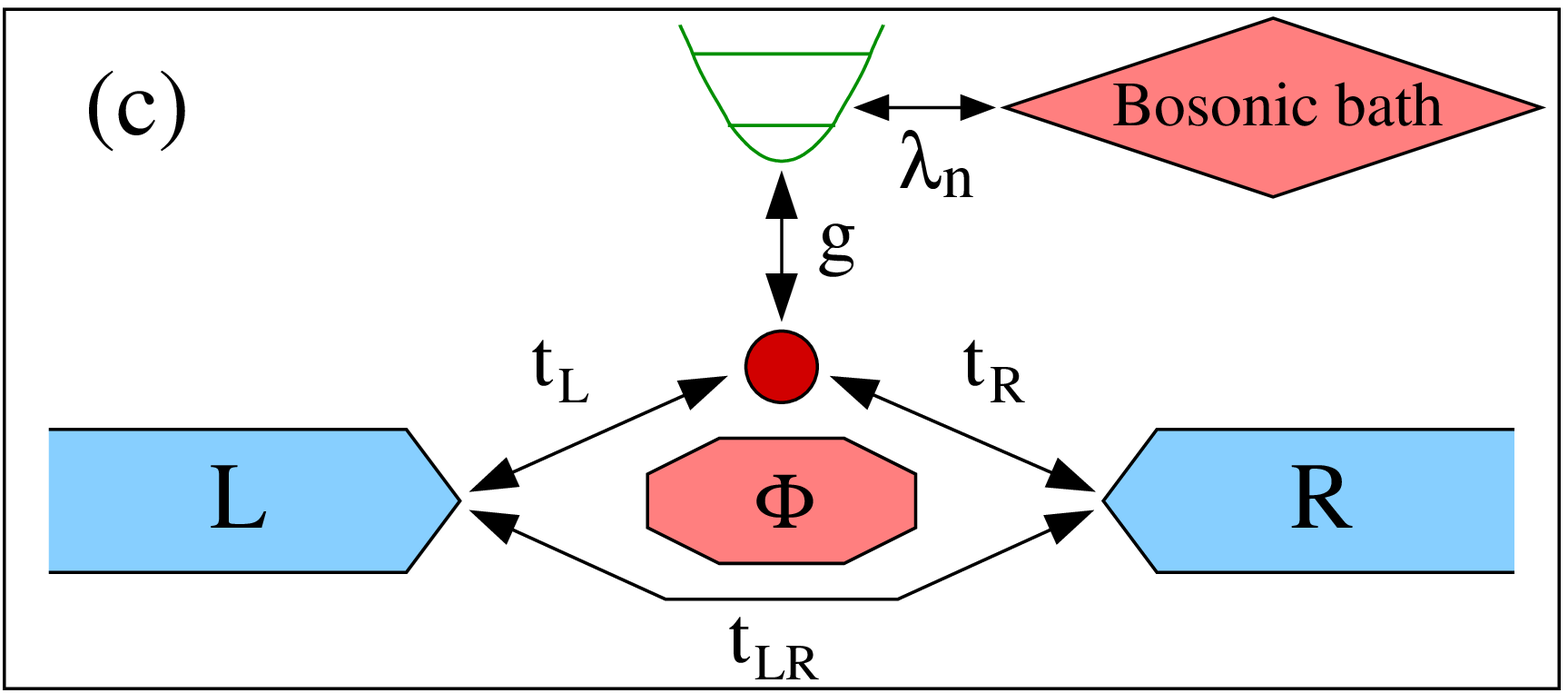}
\caption{(Color online) A schematic description of the three electronic systems we consider, and which can be mapped onto the Hamiltonian of eq.~(\ref{H}). Top to bottom: (a) a molecule adsorbed on a surface, with the bulk phonons serving as a bosonic thermal bath; (b) a single-molecule transistor, where the phonons of the substrate are coupled to local vibrational mode; (c) an Aharonov-Bohm interferometer with a molecular device embedded in one of its arms.}
\label{Fig:fig1}
\end{center}
\end{figure}

The model we consider consists of a single molecule
coupled simultaneously to a conduction band and to a
bosonic bath, which may represent, depending on the
context, either the substrate phonons or any other
continuum of bosonic degrees of freedom. A finite
temperature gradient is applied between the
electronic and bosonic baths, generating a nonzero
heat flow from the hotter bath to the colder one.
Physically this model can describe any of a number
of systems, three of which are portrayed
schematically in Fig.~\ref{Fig:fig1}: a single
molecule adsorbed on a metallic surface (upper
panel), a molecular bridge (middle panel), and a
single-molecule transistor embedded in one of the
arms of an Aharonov-Bohm interferometer (lower panel).
Under suitable conditions, to be specified below, each of these
setups can be described by the generic
continuum-limit Hamiltonian~\cite{comment-on-units}
\begin{eqnarray}
{\cal H} \!&=&\! -i v_F
                 \int_{-\infty}^{\infty}
                      \psi^{\dagger}(x) \partial_x
                      \psi(x) dx
               + \Omega_0 b^{\dagger} b
\nonumber \\
           &+& \sum_n \omega_n
                        \gamma^{\dagger}_n \gamma_n
             + (b^{\dagger} + b)
               \sum_n \lambda_n
                      (\gamma^{\dagger}_n + \gamma_n)
\nonumber \\
           &+&\! g a (b^{\dagger} + b)
                     \! :\! \psi^{\dagger}(0)
                            \psi(0) \!:
               +\, \epsilon_d a
                 \! :\! \psi^{\dagger}(0)
                        \psi(0) \!: \, ,
\label{H}
\end{eqnarray}
which is the focus of the present work. In the above
Hamiltonbian, the one-dimensional fermionic field
$\psi^{\dagger}(x)$ represents the conduction-electron
degrees of freedom, $b^{\dagger}$ denotes the molecular
vibrational mode, and $\gamma^{\dagger}_n$ are the
modes of the bosonic bath. The fermionic field obeys
canonical anticommutation relations
$\{ \psi(x), \psi^{\dagger}(y) \} = \delta(x - y)$
subject to the regularization $\delta(0) = 1/a$, where
$a$ is a suitable short-distance cutoff. In the above model we have restricted ourselves to the case where the Coulomb repulsion between the lead and the localized fermionic level is negligible, and omitted that term. The bosonic
bath is characterized by the coupling function
\begin{equation}
\Lambda(\omega) = \sum_n
                \lambda_n^2 \delta(\omega - \omega_n) ,
\end{equation}
which is assumed to have the standard power-law form
\begin{equation}
\Lambda(\omega) = 2 \pi \alpha \omega_c
            \left(
                    \frac{\omega}{\omega_c}
            \right)^s
            \theta(\omega)
            \theta(\omega_c - \omega) .
\label{J-power-law}
\end{equation}
Here, $\omega_c$ is a high-energy cutoff and $\alpha$
is a dimensionless coupling constant parameterizing
the coupling strength to the bosonic bath. The power
$s = 1$ is of particular interest as it corresponds
to an Ohmic bath.\cite{Legget1987} The parameter $g$ describes the
displacement coupling between the electrons and the
vibrational mode, while $\epsilon_d$ represents
ordinary potential scattering.

The generic model described by the Hamiltonian of Eq.~(\ref{H}) is found in discussions of different physical systems consisting of a single molecule which is coupled to large environments. It has a long history that dates
back to the 1970s, when it was proposed as a
model for the electron-phonon coupling in
mixed-valence compounds.~\cite{SvM75} In the
modern context of nanostructures it is expected
to properly describe the physics of single-molecule
devices away from Coulomb-blockade valleys where a
single unpaired spin resides on the molecule. We shall now turn to present several of these setups.

\subsection{Molecule adsorbed on a metallic surface}
The most direct realization of the model given by Eq.~(\ref{H}) is that of a single molecule adsorbed on a metallic surface. Adsorbed molecules on surfaces have been intensely studied over the years, both theoretically and experimentally (for a brief review see Ref.~[\onlinecite{cavanagh93}]).

In such molecules, the vibrational modes of adsorbed molecules relax by interaction with the surface conductance electrons as well as by coupling to the bulk phonons, and the interaction between the electrons and the vibrational mode plays an important role.~\cite{Franke2012} One of the common models used to describe such interactions~\cite{Gadzuk92} is the Hamiltonian of Eq.~(\ref{H}), prior to the reduction of the relevant conductance electrons degrees of freedom to one-dimensional fields. Adopting that description, the bosonic bath described by the operators $\gamma_n$ and $\gamma^{\dagger}_n$ in our Hamiltonian represent the bulk phonons. The molecule is then brought into contact with an STM-tip or other electronic lead, and the conductance electrons couple to the electronic level on the molecule. Focusing on the electronic mode that couples to the molecule, we can map them onto the one-dimensional field that is represented by the electronic field operators $\psi(x)$ and $\psi^{\dagger}(x)$ in the Hamiltonian. A detail of the process can be found in Ref.~[\onlinecite{VSA2012}]. In such a system, generally one would take $\epsilon_d=0$, as the conductions electrons are held in resonance with the electronic level on the molecule.

\subsection{Molecular bridge}

Another system that can be described by the model of Eq.~(\ref{H}) is that of a molecular bridge --- a single molecule sandwiched between two electronic leads. In such single-molecule devices the electron-phonon interaction plays an important role, as the molecular orbitals are coupled simultaneously to the lead electrons and to the vibrational modes of the molecule itself. The molecular bridge is typically placed on an insulating substrate which provides an additional phononic bath that couples to the molecular vibrational modes.~\cite{Lozanne93-5,Ritchie2009}

Focusing on a molecular junction held between leads with no bias voltage, one can choose a symmetric and anti-symmetric basis for the lead electrons, thus the molecule interacts with an effective single electronic band. The molecule itself is modeled by a single spinless electronic level $d^{\dagger}$ with energy $\epsilon_d$ which is coupled by displacement to a local vibrational mode $b^{\dagger}$ with frequency $\Omega_0$. This vibrational mode is further coupled by displacement to a bath of phonons $\gamma^{\dagger}_n$. The level is then coupled to the single band of electrons via a hopping matrix element $t$. The resulting Hamiltonian is given by
\begin{eqnarray}
	\mathcal{H} &=& \mathcal{H}_0 + \mathcal{H}_{\rm b} +
	\epsilon_d \hat{n}_d + \Omega_0 b^{\dagger}b +
	g\left(b^{\dagger}+b\right)\left(\hat{n}_d-\frac{1}{2}\right) +
	\nonumber \\ &&
	\left(b^{\dagger}+b\right)\sum_n \lambda_n (\gamma^{\dagger}_n+\gamma_n)
	\label{H_ele-phon_1}
\end{eqnarray}
with $\hat{n}_d = d^{\dagger}d$ and
\begin{eqnarray}
	\mathcal{H}_0 &=& \sum_k \epsilon_k c^{\dagger}_k c_k +
	t\sum_k \left(d^{\dagger}c_k + c^{\dagger}_k d\right) \nonumber \\
	\mathcal{H}_{\rm b} &=& \sum_n \omega_n \gamma^{\dagger}_n \gamma_n.
	\label{H_ele-phon_2}
\end{eqnarray}
Typically the Hamiltonian is treated either
in the weak-coupling limit using perturbation
theory in $g$, or using the Lang-Firsov
transformation~\cite{LF62} and the polaronic
approximation in the limit where $t$ is small. In the limit of a broad level close to resonance, however, this Hamiltonian can be reduced~\cite{VSA2012,Dora_Halbritter_2009} to the continuum limit Hamiltonian of Eq.~(\ref{H}) as described in detail in Ref.~[\onlinecite{VSA2012}]. This mapping process give rise to an effective high-energy cutoff which is determined by the hybridization width of the level $\Gamma=\pi\rho_0 t^2$. The effective bandwidth $D_{\rm eff}$ of the continuous field $\psi(x)$ is related to the hybridization width by $D_{\rm eff} = \pi\Gamma /2$.

\subsection{Aharonov-Bohm interferometer with a molecular device}

The final system considered is an Aharonov-Bohm interferometer with a molecular device embedded in one of its arms. This setup is a more complicated variant of the molecular bridge described above, as there is an additional transmission channel between the two leads, and a magnetic flux $\varphi$ threading the ring. Lately, this system has attracted a considerable interest, following the work of Entin-Wohlman and others~\cite{Entin2010,Entin2012} who computed the transport coefficients relating the linear-response electric and heat currents to the voltage bias and temperature gradient.

Restricting our attention to the case where the two electronic leads are held in identical temperature and chemical potential, the Hamiltonian describing such a device is given~\cite{Schoeller2001} by $\mathcal{H}=\mathcal{H}_0+\mathcal{H}_{L,R}+\mathcal{H}_{\rm dot}$, with
\begin{eqnarray}
	\mathcal{H}_0 &=& \sum_{k,a=L,R} \epsilon_{k}c^{\dagger}_{k,a} c_{k,a},
	\nonumber \\
	\mathcal{H}_{L,R} &=& W\sum_{k,q}\left(e^{-i\varphi}c^{\dagger}_{k,L}c_{q,R}+
									{\rm h.c.}\right), \nonumber \\
	\mathcal{H}_{\rm dot} &=& \sum_k\left(V_L c^{\dagger}_{k,L}d+
												V_R c^{\dagger}_{q,R}d+{\rm h.c.}\right)+
										\nonumber \\ &&
										\epsilon_d d^{\dagger}d + 
										\mathcal{H}_{\rm int}.
\end{eqnarray}
Here, $\mathcal{H}_0$ describes the left (L) and right (R) lead electrons, $\mathcal{H}_{L,R}$ describes the arm of the interferometer without the device, and $\mathcal{H}_{\rm dot}$ describes the device itself and its coupling to the leads. The general term $\mathcal{H}_{\rm int}$ may contain any local interactions at the dot that do not involve the lead electrons. In our case, it will include the local vibrational mode, the electron-phonon coupling and the bosonic bath. However, for the purpose of the mapping shown here, no further assumptions are needed on the form of $\mathcal{H}_{\rm int}$ beyond it not involving the lead electrons.

We begin by transferring to the symmetric and anti-symmetric basis, defining the new symmetric field
\begin{equation}
	\psi(x) = \frac{1}{\sqrt{V_L^2+V_R^2}}\left[V_L\psi_L(x)+V_R\psi_R(x)\right]
\end{equation}
where $\psi_{L/R}(x) = \sum_k e^{i\epsilon_k x/v_F} c_{k,L/R}$. The Green function $G_{\psi\psi^{\dagger}}(z)$ pertaining to this field at $x=0$ can now be calculated from the Green functions of the left and right fields at the origin, given by matrix form as
\begin{eqnarray}
	G(z) &=& \left(\begin{array}{cc} G_{LL}(z) & G_{LR}(z) \\
												G_{RL}(z) & G_{RR}(z)\end{array}
		\right) = \nonumber \\ &&
		\left(\begin{array}{cc} \left[g_L(z)\right]^{-1} & -te^{i\varphi} \\
												-te^{-i\varphi} &
										\left[g_R(z)\right]^{-1}\end{array}
		\right)^{-1}.
\end{eqnarray}
Using this we write the dot's Green function as
\begin{equation}
	G_{dd^{\dagger}}(z) = \left[z-\epsilon_d-\bar{V}^2G_{\psi\psi^{\dagger}}(z)
			-\Sigma_{\rm int}(z)\right]^{-1},
\end{equation}
where $\Sigma_{\rm int}(z)$ is the self-energy contribution of the $\mathcal{H}_{\rm int}$. Close to resonance, the role of the symmetric electronic Green function $G_{\psi\psi^{\dagger}}$ will be to renormalize $\epsilon_d$ and the level width. Therefore, our effective Hamiltonian of eq.~(\ref{H}) generates the same correlation functions pertaining to the dot operators as the original Hamiltonian, ensuring the validity of the calculations of the heat current. To this end, we have to adjust the parameters of our effective Hamiltonian such that the effective electronic band-width is
\begin{equation}
	D_{\rm eff} = \frac{\pi}{4} \bar{\Gamma},
\end{equation}
with $\bar{\Gamma}=(\Gamma_L+\Gamma_R)/(1+x)$, where
\begin{eqnarray}
	\Gamma_{a} &=& \pi\rho_0V_a^2, \nonumber \\
	x &=& (\pi\rho_0W)^2.
\end{eqnarray}
The renormalized level energy is effected by the flux threading the ring
\begin{equation}
	\epsilon_d \to \epsilon_d - \frac{1}{2}
					\sqrt{\alpha x}\bar{\Gamma}\cos(\varphi),
\end{equation}
where $\alpha=4\Gamma_L\Gamma_R/(\Gamma_L+\Gamma_R)^2$.

As the flux $\varphi$ contributes only to the effective $\epsilon_d$, which will be shown to effect the heat current only at quartic orders or through the renormalization of the coupling coefficient $g$, one concludes that the magnetic flux effects the heat current similarly. Moreover, its effect is an even function of the flux.

\section{System Setup and Mapping}
\label{sec:mapping}

The nonequilibrium conditions that will give rise to the heat current will be manifested by assuming to hold the electronic and bosonic baths
at different temperatures $T_{\rm e}$ and $T_{\rm b}$,
respectively. This temperature gradient may be intentional
and well controlled, or can be the by-product of some
other dynamics that inevitably causes an imbalance
between the two heat baths. In either case we assume
that all other energy relaxation mechanisms between
the electronic and the bosonic bath are sufficiently
inefficient such that all local relaxation can be
regarded as mediated by the molecule. In the Keldysh
spirit,\cite{Keldysh1965} we account for the temperature difference by
taking the initial density operator in the distant
past to have the form
\begin{equation}
\hat{\rho}_0 =
     \frac{e^{-(\beta_{\rm e} {\cal H}_{\rm e}^{0}
              +\beta_{\rm b} {\cal H}_{\rm b}^{0})}}
          {{\rm Tr} \{ e^{-(\beta_{\rm e}
                           {\cal H}_{\rm e}^{0}
                          -\beta_{\rm b}
                           {\cal H}_{\rm b}^{0})} \} } ,
\end{equation}
where $\beta_{\rm e} = 1/T_{\rm e}$ and
$\beta_{\rm b} = 1/T_{\rm b}$ are the reciprocal
temperatures and
\begin{align}
{\cal H}_{\rm e}^{0} &=
     -i v_F \int_{-\infty}^{\infty}
                 \psi^{\dagger}(x) \partial_x \psi(x) dx ,
\\
{\cal H}_{\rm b}^{0} &=
      \sum_n \omega_n \gamma^{\dagger}_n \gamma_n
\end{align}
represent the two decoupled baths. The system is then
evolved in time according to the full Hamiltonian
${\cal H}$ until steady state is reached.

The electronic Hamiltonian ${\cal H}_{\rm e}^{0}$
contains a natural high-energy cutoff or bandwidth
$D_{\rm eff} = \pi v_F/a$, which, depending on the
context, may represent either the actual
conduction-electron bandwidth or the hybridization
width of a certain molecular orbital. Our subsequent
solution of the nonequilibrium state is confined
to the weak-coupling regime, $D_{\rm eff}
\gg \max \{ g, g^2/\omega_0, |\epsilon_d| \}$,
which serves as a prerequisite for some of the
realizations of Hamiltonian of Eq.~(\ref{H})
depicted in Fig.~\ref{Fig:fig1}. In this limit
one can apply Abelian bosonization~\cite{Haldane81}
to convert the Hamiltonian and the initial density
operator to a form quadratic in bosonic operators.
Specifically, the Hamiltonian of Eq.~(\ref{H}) takes
the form
\begin{align}
{\cal H} &=
      \sum_{k > 0} \epsilon_k a^{\dagger}_k a_k
      + \sum_n \omega_n \gamma^{\dagger}_n \gamma_n
      + \Omega_0 b^{\dagger}b
\nonumber \\
     &+ \left[
                \tilde{g} (b^{\dagger} + b)
                + \tilde{\epsilon}_d
        \right]
        \sum_{k > 0} \xi_k (a^{\dagger}_k + a_k)
\nonumber \\
     &+ (b^{\dagger} + b)
        \sum_n \lambda_n
               (\gamma^{\dagger}_n + \gamma_n) ,
\label{H-bosonic}
\end{align}
where the first two terms on the right-hand
side correspond to the free Hamiltonian terms
${\cal H}_{\rm e}^{0}$ and ${\cal H}_{\rm b}^{0}$
that appear in $\hat{\rho}_0$; $a^{\dagger}_k$
and $a_k$ with $k = 2 \pi m/L$ ($L$ being the
size of the system) are canonical bosonic creation
and annihilation operators corresponding to the
Fourier modes of the fermionic density; $\epsilon_k$
equals $v_F k$; and the coefficients $\xi_k$, which
have the dimension of one over length, are given by
\begin{equation}
\xi_k = \sqrt{\frac{k}{2\pi L}}\,e^{-ak/2\pi} .
\label{xi_k}
\end{equation}
Note that we have omitted in
Eq.~(\ref{H-bosonic}) the contribution of the
$k = 0$ mode of the fermionic density as it
has no effect on our problem of interest. As
for the coupling constants $\tilde{g}$ and
$\tilde{\epsilon}_d$, these have the dimension
of energy times length, and are given to linear
order in $g$ and $\epsilon_d$ by~\cite{comment-on-units}
\begin{align}
\tilde{g} &= g a =
          \pi v_F \frac{g}{D_{\rm eff}} ,
\\
\tilde{\epsilon}_d &= \epsilon_d a =
          \pi v_F \frac{\epsilon_d}{D_{\rm eff}} .
\end{align}
As we shall argue below, one must include higher
orders in $\epsilon_d$ to account for its effect
on the heat current.

Our interest is in the steady-state heat current
flowing between the two baths. Formally there
are several heat-current operators one can define,
e.g., the heat current flowing into the bosonic
bath or the heat current flowing out of the
electronic bath, all of which must coincide in
steady state. For convenience we shall focus on
the heat current flowing into the bosonic bath,
whose corresponding operator
\begin{equation}
\hat{J}_Q = \left.
                   \frac{d {\cal H}_b^{0}(t)}{dt}
            \right|_{t = 0}
          = -i \left[
                       {\cal H}_b^{0},
                       {\cal H}
               \right]
\end{equation}
reads
\begin{equation}
\hat{J}_Q = i (b^{\dagger} + b)
            \sum_n \lambda_n \omega_n
                   (\gamma_n - \gamma^{\dagger}_n) .
\end{equation}
Our goal is to evaluate the steady-state expectation
value $J_Q = \langle \hat{J}_Q \rangle$.

\section{Resonance condition}
\label{sec:resonance_heat_current}

First we consider the case where $\tilde{\epsilon}_d = 0$,
which corresponds for a molecular bridge to electronic
resonance conditions. Technically, this limit is somewhat easier to
address as the bosonic Hamiltonian is purely quadratic.
A nonzero $\tilde{\epsilon}_d$ introduces a term
linear in bosonic operators to the Hamiltonian of
Eq.~(\ref{H-bosonic}), whose treatment requires some
care. We defer discussion of the off-resonance case
to Sec.~\ref{sec:off-resonance} below.

\subsection{Derivation of the heat current}

To compute the heat current we begin by writing it in
the form
\begin{equation}
J_Q = -2 \sum_n \lambda_n \omega_n
               {\rm Im} \{ G^{>}_{b, n}(t, t) \} ,
\label{J_Q-via-G}
\end{equation}
where
\begin{equation}
G^{>}_{b, n}(t, t') =
      \langle
              b(t) (\gamma_n - \gamma_n^{\dagger})(t')
      \rangle .
\end{equation}
In steady state all two-time correlation functions
reduce to a function of the time difference only.
This allows one to convert to the energy domain by
Fourier transforming with respect to the time
difference. We apply this procedure to
$G^{>}_{b, n}(t, t')$, and aiming at evaluating
\begin{equation}
	G^{>}_{b, n}(t, t) = \int\!\frac{d\epsilon}{2\pi} G_{b,n}^{>}(\epsilon),
\end{equation}
we resort to ordinary diagrammatic perturbation expansion to write an expression for $G_{b,n}^{>}(\epsilon)$. To this end, it is useful to define the free (i.e., $\lambda_n=0$) Green functions of the bath bosons in matrix form as
\begin{equation}
g^{r, a}_{\gamma, n}(\epsilon) =
       \left[
              \begin{array}{cc}
                    (\epsilon - \omega_n \pm i\eta)^{-1}
                    & 0
                    \\ \\
                    0 &
                  - (\epsilon + \omega_n \pm i\eta)^{-1}
              \end{array}
       \right],
\end{equation}
for the retarded and advanced functions, and
\begin{equation}
g^{<, >}_{\gamma, n}(\epsilon) =
       \pm 2 \pi n_b(\pm \epsilon)
       \left[
              \begin{array}{cc}
                    \delta (\epsilon - \omega_n) & 0
                    \\ \\
                    0 & -\delta(\epsilon + \omega_n)
              \end{array}
       \right],
\end{equation}
for the lesser and greater functions, where $n_b(\epsilon) = 1/(e^{\beta_b \epsilon} - 1)$ is
the Bose-Einstein distribution function corresponding
to the temperature $T_b$. Similarly, we shall denote the fully dressed Green functions of the molecular vibrational mode $b^{\dagger}$ by $G^{r, a}(\epsilon)$ and
$G^{<, >}(\epsilon)$, for which the $2\!\times\!2$ matrix notation
\begin{equation}
G^{\nu}(\epsilon) =
   \left[
           \begin{array}{cc}
                G^{\nu}_{b b^{\dagger}}(\epsilon) &

                G^{\nu}_{b b}(\epsilon)
                \\ \\
                G^{\nu}_{b^{\dagger} b^{\dagger}}(\epsilon)
                & G^{\nu}_{b^{\dagger} b}(\epsilon)
           \end{array}
   \right]
\end{equation}
is used. [The same $2\!\times\!2$ matrix notation
applies to $g^{\nu}_{\gamma, n}(\epsilon)$ above.]

Having laid out the building blocks for the perturbation expansion, we rely on Langreth theorem~\cite{Langreth76} to have the identity, at steady-state
\begin{widetext}
\begin{equation}
{\lim_{t\to\infty}}G^{>}_{b, n}(t, t) =
    \int_{-\infty}^{\infty}
         \frac{d \epsilon}{2 \pi}
         \lambda_n
         \sum_{p = 1, 2}
         \left[
                  \left[
                          G^{>}(\epsilon)
                  \right]_{1 p}
                  \left\{
                           \left[
                                  g^a_{\gamma, n}(\epsilon)
                           \right]_{22}
                         - \left[
                                  g^a_{\gamma, n}(\epsilon)
                           \right]_{11}
                  \right\}
                + \left[
                          G^{r}(\epsilon)
                  \right]_{1 p}
                  \left\{
                           \left[
                                  g^>_{\gamma, n}(\epsilon)
                           \right]_{22}
                         - \left[
                                  g^>_{\gamma, n}(\epsilon)
                           \right]_{11}
                  \right\}
         \right],
\label{G^>_b,n}
\end{equation}
\end{widetext}
which we shall now turn to evaluate.

Focusing initially on the retarded and advanced Green
functions $G^{r, a}(\epsilon)$, these acquire the form
\begin{equation}
G^{r, a}(\epsilon) =
   \left[
           \begin{array}{cc}
                \epsilon - \Omega_0
                         - \Sigma^{r, a}(\epsilon) &
                - \Sigma^{r, a}(\epsilon)
                \\ \\
                - \Sigma^{r, a}(\epsilon) &
                - \epsilon - \Omega_0
                           - \Sigma^{r, a}(\epsilon)
           \end{array}
   \right]^{-1} ,
\label{G^ra}
\end{equation}
where
\begin{align}
\Sigma^{r, a}(\epsilon) &=
       \tilde{g}^2
       \sum_{k > 0} \xi_k^2
            \left[
                    \frac{1}
                         {\epsilon - \epsilon_k \pm i\eta}
                  - \frac{1}
                         {\epsilon + \epsilon_k \pm i\eta}
            \right]
\nonumber \\
   & + \sum_n \lambda_n^2
            \left[
                    \frac{1}
                         {\epsilon - \omega_n \pm i\eta}
                  - \frac{1}
                         {\epsilon + \omega_n \pm i\eta}
            \right] ,
\label{S^ra}
\end{align}
are the corresponding self-energy functions. There are
two contributions to $\Sigma^{r, a}(\epsilon)$: one due
to the coupling to the electronic bath [the first term
on the right-hand side of Eq.~(\ref{S^ra})], and another
due to the coupling to the bosonic bath [the second term
on the right-hand side of Eq.~(\ref{S^ra})]. Denoting
these two terms by $\Sigma^{r, a}_{\, e}(\epsilon)$ and
$\Sigma^{r, a}_{\, b}(\epsilon)$, respectively, the
former can be expressed in a closed analytical
form~\cite{VSA2012} in terms of the Exponential
Integral function:~\cite{AS-E1}
\begin{align}
\Sigma^{r, a}_{\, e}(\epsilon) =
            (\rho_0 \tilde{g})^2 D_{\rm eff}
          & \left[
                    \xi e^{\xi}
                    E_1 (\xi \pm i\eta)
            \right.
\nonumber \\
          &\,\, \left.
                    - \xi e^{-\xi}
                      E_1 (-\xi \mp i\eta)
                    - 2
            \right] .
\label{S^ra_e}
\end{align}
Here, $\rho_0 = 1/(2 \pi v_F) = 1/(2 a D_{\rm eff})$
is the density of states per unit length, and $\xi$
equals $\epsilon/D_{\rm eff}$. As for the second
contribution $\Sigma^{r, a}_{\, b}(\epsilon)$, it
can be conveniently expressed in terms of the
coupling function $\Lambda(\omega)$,
\begin{equation}
\Sigma^{r, a}_{\, b}(\epsilon) =
       \int_{0}^{\infty}
            \Lambda(\epsilon')
            \left[
                    \frac{1}
                         {\epsilon - \epsilon' \pm i\eta}
                  - \frac{1}
                         {\epsilon + \epsilon' \pm i\eta}
            \right] d\epsilon' .
\label{S^ra_b}
\end{equation}
For an Ohmic bath with $s = 1$, Eq.~(\ref{S^ra_b})
can be evaluated in closed analytic form to obtain
\begin{equation}
\Sigma^{r, a}_{\, b}(\epsilon) =
       2\pi \alpha\!
       \left[
               \epsilon
               \ln \left|
                           \frac{\epsilon + \omega_c}
                                {\epsilon - \omega_c}
                   \right|
               - 2 \omega_c
               \mp i \pi \epsilon\,
                   \theta \!\left(
                                    \omega_c^2 - \epsilon^2
                            \right)
       \right] .
\label{S^ra_b-Ohmic}
\end{equation}

Proceeding to the lesser and greater Green functions
$G^{<, >}(\epsilon)$, these read
\begin{equation}
G^{<, >}(\epsilon) =
        G^{r}(\epsilon)
        \left[
               \begin{array}{cc}
               1 & 1 \\
               1 & 1
               \end{array}
        \right]
        \Sigma^{<, >}(\epsilon)
        G^{a}(\epsilon) ,
\label{G^<>}
\end{equation}
with the lesser and greater self-energy functions
\begin{align}
\Sigma^{<, >}(\epsilon) = &
       \pm 2\pi n_{e}(\pm \epsilon)
           (\rho_0 \tilde{g})^2
           \epsilon\, e^{-|\epsilon|/D_{\rm eff}}
\nonumber \\
     & \pm 2\pi n_{b}(\pm \epsilon)
           \left[
                   \Lambda(\epsilon) - \Lambda(-\epsilon)
           \right] .
\label{S^<>}
\end{align}
Here, $n_e(\epsilon) = 1/(e^{\beta_e \epsilon} - 1)$ is
the Bose-Einstein distribution function corresponding
to the temperature $T_e$. Note that Eq.~(\ref{S^<>})
can be conveniently related to the two components of
the retarded self-energy through
\begin{equation}
\Sigma^{<, >}(\epsilon) =
       \mp 2 n_{e}(\pm \epsilon)
               {\rm Im} \{ \Sigma^{r}_{e}(\epsilon) \}
       \mp 2 n_{b}(\pm \epsilon)
               {\rm Im} \{ \Sigma^{r}_{b}(\epsilon) \} ,
\label{S^<>-via-S^r}
\end{equation}
which generalizes the standard equilibrium relation
$\Sigma^{<, >}(\epsilon) = \mp 2 n(\pm \epsilon)
{\rm Im} \{ \Sigma^{r}_{e}(\epsilon) \}$.

To evaluate $G^{>}_{b, n}(t, t)$ of
Eq.~(\ref{G^>_b,n}), it is useful to utilize the
identities
\begin{align}
\sum_{p = 1, 2}
     \left[
             G^{r}(\epsilon)
     \right]_{1 p} &=
     |g(\epsilon)|^2 2 \Omega_0 ( \epsilon + \Omega_0 )
                       \Sigma^{>}(\epsilon) ,
\nonumber \\
\sum_{p = 1, 2}
     \left[
             G^{r}(\epsilon)
     \right]_{1 p} &=
     |g(\epsilon)|^2 ( \epsilon + \Omega_0 )
     \left[
             \epsilon^2 - \Omega_0^2
             - 2\Omega_0 \Sigma^{a}(\epsilon)
     \right] ,
\end{align}
which follow directly from Eqs.~(\ref{G^ra}) and
(\ref{G^<>}). Here we have introduced the auxiliary
function
\begin{equation}
g(\epsilon) = \frac{1}
                   {\epsilon^2 - \Omega_0^2
                    - 2\Omega_0 \Sigma^{r}(\epsilon)} ,
\label{g-def}
\end{equation}
and made use of the fact that $\Sigma^{a}(\epsilon)
= \left[ \Sigma^{r}(\epsilon) \right]^{\ast}$.
Inserting these identities into Eq.~(\ref{G^>_b,n})
one obtains
\begin{widetext}
\begin{equation}
-{\rm Im} \{\lim_{t\to\infty} G^{>}_{b, n}(t, t) \} =
      \lambda_n
      \int_{-\infty}^{\infty}
           d \epsilon
           |g(\epsilon)|^2 \Omega_0 (\epsilon + \Omega_0 )
           \left[
                   \Sigma^{>}(\epsilon)
                   - 2 {\rm Im} \{ \Sigma^{r}(\epsilon) \}
                     n_b(-\epsilon)
           \right]
           \left[
                   \delta(\epsilon - \omega_n)
                   + \delta(\epsilon + \omega_n)
           \right] ,
\end{equation}
which reduces by virtue of Eq.~(\ref{S^<>-via-S^r}) to
\begin{equation}
-{\rm Im} \{\lim_{t\to\infty} G^{>}_{b, n}(t, t) \} =
      \lambda_n
      \int_{-\infty}^{\infty}
           d \epsilon
           |g(\epsilon)|^2 \Omega_0 (\epsilon + \Omega_0 )
           2 {\rm Im} \{ \Sigma^{r}_{e}(\epsilon) \}
           \left[
                   n_e(-\epsilon) - n_b(-\epsilon)
           \right]
           \left[
                   \delta(\epsilon - \omega_n)
                   + \delta(\epsilon + \omega_n)
           \right] .
\end{equation}
Plugging this result into Eq.~(\ref{J_Q-via-G}),
employing the relations
$g(-\epsilon) = g^{\ast}(\epsilon)$ and
$[ n_e(-\epsilon) - n_b (-\epsilon) ] =
- [ n_e(\epsilon) - n_b (\epsilon) ]$, and exploiting
the fact that $\Lambda(\epsilon)$ is restricted to positive
energies, we finally arrive at
\begin{equation}
J_Q = (4 \pi \rho_0 \tilde{g} \Omega_0)^2
      \int_{0}^{\infty}
           \frac{d \epsilon}{2 \pi}
           \frac{\epsilon^2 e^{-\epsilon/D_{\rm eff}}}
                {| \epsilon^2 - \Omega_0^2
                   - 2\Omega_0 \Sigma^{r}(\epsilon) |^2}
           \Lambda(\epsilon)
           \left[
                   n_e(\epsilon) - n_b(\epsilon)
           \right] ,
\label{J_Q-general}
\end{equation}
\end{widetext}
where we have explicitly written out the function
$|g(\epsilon)|^2$ that appears in the integrand
[see Eq.~(\ref{g-def})].

Equation~(\ref{J_Q-general}) is the central result of
this paper. It provides an exact expression for the
heat current corresponding to the bosonic Hamiltonian of
Eq.~(\ref{H-bosonic}), for a general coupling function
$\Lambda(\epsilon)$. Since $[n_e(\epsilon) - n_b(\epsilon)]$
with $\epsilon > 0$ is positive definite for
$T_e > T_b$ (negative definite for $T_b > T_e$), the
heat current flows, as it physically should, from the
hotter bath to the colder one. Below we analyze in
detail the characteristics of $J_Q$ in different
temperature and coupling regimes.

The form of expression for the heat current bears a similarity to Landauer formula. Other works studying thermal conductance through local contacts have arrived at a similar expressions or used a Landauer-type expression as a starting point.~\cite{Segal2003,Ozpineci2001,Wang2008,Segal2011} In the context of heat current between bosonic reservoirs this expression was derived in earlier works discussing Hamiltonians similar to the one in eq.~(\ref{H-bosonic})~\cite{Segal2008,Dahr2008,Ojanen2008}. This type of expression survives even when accounting perturbatively for the interaction with vibrational modes.~\cite{Galperin2007}

\subsection{Low-temperature limit}

We begin with the low-temperature limit,
$T_e, T_b \ll \Omega_0$ (throughout this paper we
assume that $\Omega_0 < \omega_c, D_{\rm eff}$). In
this limit, the Bose-Einstein distribution functions
$n_e(\epsilon)$ and $n_b(\epsilon)$ that enter the
integrand of Eq.~(\ref{J_Q-general}) have decayed long
before $e^{-\epsilon/D_{\rm eff}}$ and $|g(\epsilon)|^2$
have changed in any significant manner from their
$\epsilon = 0$ values. Thus, to a good approximation
one can
(i) set $e^{-\epsilon/D_{\rm eff}} |g(\epsilon)|^2
    \to |g(0)|^2$ in the integrand of
    Eq.~(\ref{J_Q-general}), and
(ii) extend the upper integration limit to infinity.
Taking the coupling function $\Lambda(\epsilon)$ to have
the power-law form of Eq.~(\ref{J-power-law}) this
yields
\begin{equation}
J_Q \approx (4 \pi \rho_0 \tilde{g} \Omega_0)^2
      \alpha \omega_c^{1 - s} |g(0)|^2 \!
      \int_{0}^{\infty}\!\!
           \epsilon^{2 + s}
           \left[
                   n_e(\epsilon)\!-\!n_b(\epsilon)
           \right] d \epsilon .
\end{equation}
The resulting integral can now be carried out
analytically using
\begin{equation}
\int_{0}^{\infty}\!\!
     \frac{ \epsilon^{2 + s} }
          { e^{\beta \epsilon} - 1 } d \epsilon
     = \beta^{-(3 + s)} \Gamma(3 + s) \zeta(3 + s) ,
\end{equation}
where $\zeta(x)$ is the Riemann zeta
function.~\cite{AS-zeta} This in turn gives
\begin{equation}
J_Q \approx A ( T_e^{3 + s} - T_b^{3 + s} ) ,
\label{J_Q-low-T}
\end{equation}
with $A = (4 \pi \rho_0 \tilde{g} \Omega_0)^2
\alpha \omega_c^{1 - s} |g(0)|^2 \Gamma(3 + s)
\zeta(3 + s)$. Lastly, $g(0)$ has the explicit
expression
\begin{equation}
g(0) = \frac{1}{-\Omega_0^2 + 4 \Omega_0
                 \left[
                         D_{\rm eff} (\rho_0 \tilde{g})^2
                         + 2\pi \alpha \omega_c/s
                 \right]} ,
\end{equation}
allowing one to express the coefficient $A$ entirely in
terms of the basic model parameters entering the bosonic
Hamiltonian of Eq.~(\ref{H-bosonic}).

As can be seen from Eq.~(\ref{J_Q-low-T}), the
low-temperature heat current shows a rather strong
temperature dependence. In particular, the
linear-response heat conductance
\begin{equation}
\sigma_Q = \lim_{\Delta T \to 0} \frac{1}{\Delta T}
              J_Q(T_e = T + \Delta T, T_b = T)
\label{conductance_defined}
\end{equation}
varies as $T^{2 + s}$, corresponding to $T^3$ for an
Ohmic bath. This should be contrasted with the heat
conductance of a generic noninteracting electronic
tunnel junction, which varies linearly with $T$ at
sufficiently low temperature.~\cite{Hanggi2011} For the heat conductance
of a tunnel junction to display a superlinear
temperature dependence of the form found here, its
transmission coefficient must vanish in a power-law
fashion at the Fermi energy.

This power-law behavior of the heat current at low temperatures was also observed for systems in which the two bosonic reservoirs are connected by a system with few degrees of freedom such as in the spin-boson model~\cite{Ojanen2011,Saito2013}. The similarity between these systems and the system under consideration here, at low-temperatures with respect to $\Omega_0$, stems from the fact that at this temperature regime only the lowest lying levels of the vibrational mode of the molecule are available for transferring energy between the baths.

\subsection{Ohmic bath}

As commented above, an Ohmic bath with $s = 1$ is of
particular interest. Focusing on this case and on the
hierarchy $\Omega_0, T_e, T_b \ll \min \{D_{\rm eff},
\omega_c\}$, we devise below an analytical expression for the nonequilibrium
heat current, encompassing the crossover from the
low-temperature regime, $\max \{T_e, T_b\} \ll \Omega_0$,
to the intermediate-temperature one,
$\Omega_0 < \max \{T_e, T_b \}$. This expression is approximate as it employs a power series expansion of the self-energies, but yields accurate results within the regime where $D_{\rm eff}$ and $\omega_c$ are the largest energy scales in the system.

To this end, consider the function $g(\epsilon)$ which
enters the integrand of Eq.~(\ref{J_Q-general}). Since
the Bose-Einstein distribution functions $n_e(\epsilon)$
and $n_b(\epsilon)$ decay on a scale far smaller than
$\min \{ D_{\rm eff}, \omega_c \}$, it suffices to
accurately represent $g(\epsilon)$ for
$\epsilon \ll \min \{ D_{\rm eff}, \omega_c \}$.
This allows one to expand the exact expressions for
$\Sigma^{r}_{e}(\epsilon)$ and $\Sigma^{r}_{b}(\epsilon)$
[see Eqs.~(\ref{S^ra_e}) and (\ref{S^ra_b-Ohmic}) above]
in $x = \epsilon/D_{\rm eff}$ and $y =\epsilon/\omega_c$
to
obtain
\begin{align}
\Sigma^{r}_{e}(\epsilon) &=
     - (\rho_0 \tilde{g})^2 D_{\rm eff}
       \left[
              i \pi x + 2
              + {\cal O}
                \left(
                        x^2 \ln x
                \right)
       \right] ,
\\
\Sigma^{r}_{b}(\epsilon) &=
     - 2 \pi \alpha \omega_c
       \left[
              i\pi y + 2
              + {\cal O} \left(
                                 y^2
                         \right)
       \right] .
\end{align}
Settling with linear orders
in $x$ and $y$, the function $g(\epsilon)$ is well
approximated for $\epsilon \ll \min \{ D_{\rm eff},
\omega_c \}$ by
\begin{equation}
g(\epsilon) =
     \frac{1}{(\epsilon - z_{+})(\epsilon - z_{-})} \, ,
\end{equation}
where $z_{\pm}$ equals $\pm \tilde{\Omega} - i/\tau$ with the
softened frequency
\begin{equation}
\tilde{\Omega} = \Omega_0
     \sqrt{1 - 4\frac{(\rho_0\tilde{g})^2 D_{\rm eff} +
                      2\pi\alpha \omega_c}{\Omega_0}
             - \pi^2 \left[
                             (\rho_0 \tilde{g})^2
                             + 2\pi\alpha
                     \right]^2
          }
\label{omega}
\end{equation}
and the relaxation rate
\begin{equation}
\frac{1}{\tau} =
     \pi \Omega_0 \left[
                          (\rho_0 \tilde{g})^2
                          + 2\pi \alpha
                  \right] .
\label{tau}
\end{equation}
The softened frequency $\tilde{\Omega}$ characterizes the dressed excitations of the phonon, and thus serves as the energy scale which determines the cross over from the high-temperature to the low-temperature behavior. Within this approximation for $g(\epsilon)$, the heat
current for an Ohmic bath becomes
\begin{equation}
J_Q = (4 \pi \rho_0 \tilde{g} \Omega_0)^2 \alpha
      \int_{0}^{\infty}
           d \epsilon \epsilon^3
           \frac{\left[
                         n_e(\epsilon) - n_b(\epsilon)
                 \right]}
                {| (\epsilon - z_{+})
                   (\epsilon - z_{-}) |^2} ,
\label{J_Q-Ohmic-int}
\end{equation}
where we have set in addition
$e^{-\epsilon/D_{\rm eff}} \to 1$ and extended
the upper integration limit to infinity (both
approximations being well justified by the
hierarchy $T_e, T_b \ll \min \{D_{\rm eff},
\omega_c \}$).
The resulting integral in Eq.~(\ref{J_Q-Ohmic-int})
can be performed in a closed analytic form in terms
of the digamma function~\cite{AS-psi} $\psi(z)$.
Skipping details of the algebra we quote here only
the end result:
\begin{widetext}
\begin{equation}
J_Q = (2\pi \rho_0 \tilde{g} \Omega_0)^2 \alpha\,
      {\rm Im}
      \left\{
               (\tau - i/\tilde{\Omega})^2
               \left[
                      \psi
                      \left(
                              \frac{\tau^{-1} + i\tilde{\Omega}}
                                   {2\pi T_{b}}
                      \right)
                    - \psi
                      \left(
                              \frac{\tau^{-1} + i\tilde{\Omega}}
                                   {2\pi T_{e}}
                      \right)
                    - \ln \left(
                                  \frac{T_{e}}{T_{b}}
                          \right)
               \right]
               + i\pi\tau \left(
                                  T_{e} - T_{b}
                          \right)
      \right\} .
\label{J_Q-Ohmic}
\end{equation}
\end{widetext}

It is straightforward to confirm using the asymptotic
expansion~\cite{AS-psi}
\begin{equation}
\psi(z) = \ln(z) - \frac{1}{2z} - \frac{1}{12 z^2}
        + \frac{1}{120 z^4} + O(z^{-6})
\end{equation}
that Eq.~(\ref{J_Q-Ohmic}) properly reduces for
$T_e, T_b \ll \tilde{\Omega}$ to Eq.~(\ref{J_Q-low-T}) with
$s = 1$, including the precise value of the prefactor
$A$. More interesting is the limit $\tilde{\Omega} < \max
\{ T_{e}, T_{b} \}$, when the leading contribution
to Eq.~(\ref{J_Q-Ohmic}) crosses over to a linear
dependence on the temperature difference
$\Delta T = (T_{e} - T_{b})$. Thus, as the larger of
the two temperatures exceeds the vibrational resonance
energy, the heat current continues to increase
linearly with $\Delta T$. This behavior markedly
differs from that of a resonant electronic tunnel
junction, whose heat current depends logarithmically
on $\Delta T$ above the resonance energy. The physical
difference stems, as we show below, from the bosonic
nature of the vibrational mode, which can be excited to
exceedingly high energies by creating ever more phonons.
This should be contrasted with a resonant electronic
level, which can only be empty or occupied.

\subsection{Numerical results}

Having analyzed analytically certain limits, we now
proceed to a complete numerical evaluation of the
exact heat current of Eq.~(\ref{J_Q-general}) at
arbitrary temperatures $T_e$ and $T_b$. As the temperatures $T_e$ and $T_b$ enter the expression for the heat current only through the term $[n_e(\epsilon)-n_b(\epsilon)]$ in the integrand, it is anti-symmetric under replacing them. As such, we may restrict ourselves to calculations in which $T_b$ is held constant and the heat current is calculated for different values of $T_e$, and generalize the results for opposite values by inverting the direction of the current.

\begin{figure}[tb]
\begin{center}
\includegraphics[width=0.44\textwidth]{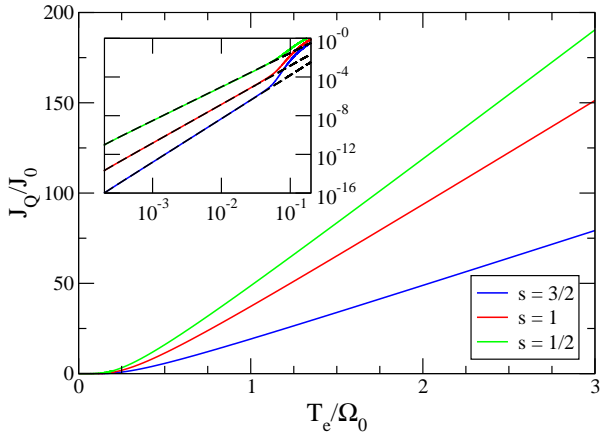}
\caption{(Color online) The heat current $J_Q$ between the electronic and bosonic baths, as a function of the temperature difference between the baths. Here the temperature of the bosonic bath was held constant at a low value $T_b/\Omega_0=10^{-4}$ and the temperature of the electronic bath $T_e$ was changed. The heat current was calculated for different types of bosonic baths, where we considered a sub-Ohmic case ($s=1/2$, green), an Ohmic case (red) and a super-Ohmic case ($s=3/2$, blue). Inset: a log-log plot of the low-temperature regime. The dashed lines are following the appropriate power-laws predicted in that regime by Eq.~(\ref{J_Q-low-T}). Here $J_Q$ is measured in units of the basic heat current $J_0 = (4\pi \rho_0 \tilde{g}\Omega_0)^2\alpha$, where $2\pi\rho_0 \tilde{g} = 0.1$, and $\alpha = 10^{-3}$ are in the weak coupling regime. The cutoffs used are $D_{\rm eff}/\Omega_0=\omega_c/\Omega_0 =20$.
		}
\label{Fig:J_Q-low-T}
\end{center}
\end{figure}

\begin{figure}[tb]
\begin{center}
\includegraphics[width=0.44\textwidth]{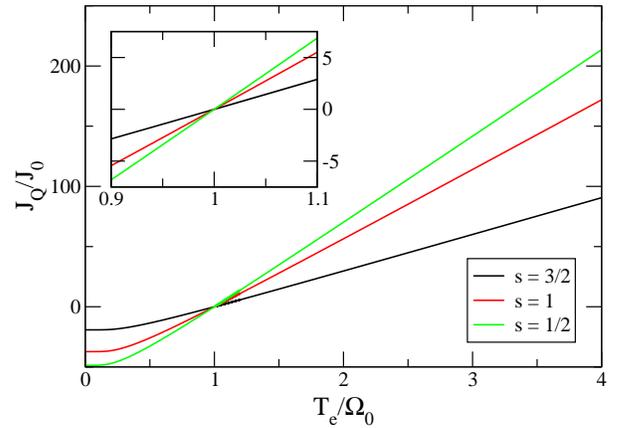}
\caption{(Color online) The heat current between the electronic and bosonic baths as a function of the temperature difference between the baths. Here the temperature of the bosonic bath was held constant at $T_b/\Omega_0 = 1$ and we calculated the current for different values of $T_e$. The heat current was calculated for different types of bosonic baths, where we have considered a sub-Ohmic case ($s=1/2$, green), an Ohmic case (red) and a super-Ohmic case ($s=3/2$, blue). Inset: a zoom over the regime about the point $T_e = T_b = \Omega_0$ where the heat current displays a linear dependence on $\Delta T$. Here $J_Q$ is measured in units of the basic heat current $J_0 = (4\pi \rho_0 \tilde{g}\Omega_0)^2\alpha$, where $2\pi\rho_0 \tilde{g} = 0.1$, and $\alpha = 10^{-3}$ are in the weak coupling regime. The cutoffs used are $D_{\rm eff}/\Omega_0=\omega_c/\Omega_0 =20$.
        }
\label{Fig:J_Q-high-T}
\end{center}
\end{figure}

In Fig.~\ref{Fig:J_Q-low-T} we have addressed the case where one bath (the bosonic one) is held at a constant low temperature $T_b/\Omega_0=10^{-4}$, while scanning different values of $T_e$ higher than that temperature. We have plotted the heat current in the case of a sub-Ohmic bath (with $s=1/2$), an Ohmic bath and a super-Ohmic bath (with $s=1$). The graph shows a clear crossover from a power-law behavior to a linear dependence as $\Delta T$ increases to values of the order of $\Omega_0$. In the inset of the graph we have plotted, on a log-log scale, the heat current for small values of $\Delta T$ for each type of bath, and a dashed line following the expected power-law behaviors at low temperatures given at Eq.~(\ref{J_Q-low-T}). The agreement between the calculated heat current and the expected one is excellent while $T_e \ll \Omega_0$, breaking at about $T_e/\Omega_0 \simeq 0.1$. Figure~\ref{Fig:J_Q-high-T} describes the heat current as well, but focuses on the case where the temperature of the bosonic bath is held at the high value of $T_b=\Omega_0$. In this case the heat current about $\Delta T = 0$ displays a linear dependence on $\Delta T$, with different slope for every type of bosonic bath.

\begin{figure}[tb]
\begin{center}
\includegraphics[width=0.44\textwidth]{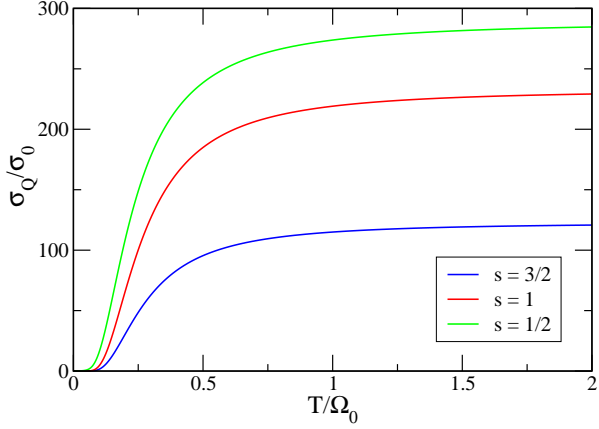}
\caption{(Color online) The linear-response heat conductance $\sigma_Q$ between the baths as a function of the temperature $T$ in which both bath are helds. The conductance is calculated for different types of bosonic baths, where we have considered a sub-Ohmic case ($s=1/2$, green), an Ohmic case (red) and a super-Ohmic case ($s=3/2$, blue). Here $\sigma_Q$ is measured in units of a basic heat conductance $\sigma_0 = (4\pi\rho_0\tilde{g})^2\Omega_0\alpha$, where $2\pi\rho_0 \tilde{g} = 0.1$, and $\alpha = 10^{-3}$ are in the weak coupling regime. The cutoffs used are $D_{\rm eff}/\Omega_0=\omega_c/\Omega_0 =20$.
        }
\label{Fig:diff_conductance}
\end{center}
\end{figure}

In Fig.~\ref{Fig:diff_conductance} we have plotted the linear-response heat conductance $\sigma_Q$, as defined in Eq.~(\ref{conductance_defined}), for different temperatures and types of baths. As the temperatures rises, the conductance increases until it saturates at about $T\simeq \Omega_0$. The different exponents relating to the type of bosonic baths are evident in the low-temperature regime.

\section{Master equation approach}
\label{sec:master_equation}

Although our solution for $J_Q$ is formally exact
within the bosonic Hamiltonian of Eq.~(\ref{H-bosonic}),
it is instructive to develop a more transparent
physical picture that would, in particular, elucidate
the source of distinction between the bosonic system
under consideration and a conventional resonant
tunnel junction. To this end, we devise below a
Master equation approach, applicable at weak coupling. This approach to address heat flow was previously applied by Segal~\cite{Segal2006,Segal2008}, considering a different setup. Leijnse and others~\cite{Flesnberg2010} have also used this method to calculate the thermopower properties in a similar setup.

The basic components of the theory are the probabilities
$P_{n}(t)$ to find the local boson $b^{\dagger}$ at
time $t$ in the state where $b^{\dagger} b = n$. These
probabilities are connected at the golden-rule
approximation by rate equations of the form
\begin{align}
\frac{d P_n}{d t} &= P_{n + 1} W_{n + 1 \to n}
                   + P_{n - 1} W_{n - 1 \to n}
\nonumber \\
                  &- P_n \left(
                                 W_{n \to n + 1}
                                 + W_{n \to n - 1}
                         \right)
\label{Master-equations}
\end{align}
(terms with $n - 1$ should be omitted for $n = 0$),
with the transition rates~\cite{comment-on-units}
\begin{align}
W_{n \to n + 1} &= 2\pi (n + 1)
     \left[
              F(\Omega_0) n_e(\Omega_0)
              + \Lambda(\Omega_0) n_b(\Omega_0)
     \right] ,
\\
W_{n \to n - 1} &= 2\pi n
     \left\{
              F(\Omega_0) [ 1 + n_e(\Omega_0) ]
     \right.
\nonumber \\
   & \;\;\;\;\;\;\;\;\;\;\;\;
     \left.
              + \Lambda(\Omega_0) [ 1 + n_b(\Omega_0) ]
     \right\} .
\end{align}
Here, $F(\Omega_0)$ equals $(\rho_0 \tilde{g})^2 \Omega_0
\exp(-\Omega_0/D_{\rm eff})$. The golden-rule approximation used here corresponds to lowest order perturbation theory in both the coupling to the phononic bath [represented by the coupling function $F(\epsilon)$] and the coupling to the bosonic bath [represented by the coupling function $\Lambda(\epsilon)$].

\subsection{Effective temperature}

It is useful to define at this point the reduced
transition rates
\begin{align}
w_{\uparrow} &= 2\pi
     \left[
              F(\Omega_0) n_e(\Omega_0)
              + \Lambda(\Omega_0) n_b(\Omega_0)
     \right] ,
\\
w_{\downarrow} &= 2\pi
     \left\{
              F(\Omega_0) [ 1 + n_e(\Omega_0) ]
              + \Lambda(\Omega_0) [ 1 + n_b(\Omega_0) ]
     \right\} ,
\end{align}
such that $W_{n \to n + 1} = (n + 1) w_{\uparrow}$ and
$W_{n \to n - 1} = n w_{\downarrow}$. Focusing on steady
state, when $d P_n/dt = 0$, Eq.~(\ref{Master-equations})
can be recast as an infinite set of coupled linear
equations,
\begin{align}
&\;\;\;\;\;\;\;\;\;\;\;\;\;\;\;\;
P_1 = (w_{\uparrow}/w_{\downarrow}) P_0 ,
\\
P_{n + 1} &=
     \left(
             \frac{n}{n+1}
             + \frac{w_{\uparrow}}{w_{\downarrow}}
     \right) P_n
   - \frac{n}{n+1}
     \frac{w_{\uparrow}}{w_{\downarrow}}
     P_{n - 1} ,
\end{align}
whose solution is $P_n = B p^n$ with
$p = w_{\uparrow}/w_{\downarrow} < 1$.
Here, $B = (1 - p)$ is a normalization factor which
comes to ensure that $\sum_{n = 0}^{\infty} P_n = 1$.
Thus, the probabilities $P_n$ obey a Boltzmann-like form
with an effective temperature $T_{\rm eff}$ defined by
\begin{equation}
p = e^{-\Omega_0/T_{\rm eff}} .
\end{equation}

Consider first the case of thermal equilibrium, when
$T_e = T_b = T$ and $n_{e}(\epsilon) = n_{b}(\epsilon)
= n(\epsilon)$. Under these circumstances
\begin{equation}
p = \frac{w_{\uparrow}}{w_{\downarrow}}
  = \frac{n(\Omega_0)}{1 + n(\Omega_0)}
  = e^{-\Omega_0/T} ,
\end{equation}
hence $T_{\rm eff}$ equals $T$ irrespective of details
of the two baths. Once a temperature gradient is applied
between the two reservoirs, i.e., $T_e \neq T_b$, then
$T_{\rm eff}$ falls in between $\min \{T_e, T_b\}$ and
$\max \{T_e, T_b\}$, as follows from the equality
\begin{equation}
e^{-\Omega_0/T_{\rm eff}} = q e^{-\Omega_0/T_{e}}
                          + (1 - q) e^{-\Omega_0/T_{b}}
\label{T_eff-via-exponents}
\end{equation}
with
\begin{equation}
q = \frac{F(\Omega_0) [ 1 + n_e(\Omega_0) ]}
         {F(\Omega_0) [ 1 + n_e(\Omega_0) ]
          + \Lambda(\Omega_0) [ 1 + n_b(\Omega_0) ]} .
\label{q-def}
\end{equation}

Equation~(\ref{T_eff-via-exponents}) can be
solved analytically in the high-temperature limit,
$\Omega_0 \ll T_e, T_b$, where Fermi's golden rule
(and thus our Master equation approach) is expected
to apply. Specifically, expanding each of the
exponents to linear order in $\Omega_0$ one obtains
\begin{equation}
T_{\rm eff} \simeq \frac{T_{e} T_{b}}
                        {q T_{b} + (1 - q) T_{e}} .
\label{T_eff-via-q}
\end{equation}
This expression can further be simplified by noting
that $q$ depends implicitly on $T_e$ and $T_b$
through the Bose-Einstein distribution functions
$n_{e}(\Omega_0) \simeq T_e/\Omega_0$ and
$n_{b}(\Omega_0) \simeq T_b/\Omega_0$. Plugging these
relations into Eq.~(\ref{q-def}) and inserting the
resulting expression for $q$ into
Eq.~(\ref{T_eff-via-q}), one finally arrives at
\begin{equation}
T_{\rm eff} = \frac{F(\Omega_0)}
                   {F(\Omega_0) + \Lambda(\Omega_0)} T_e
            + \frac{\Lambda(\Omega_0)}
                   {F(\Omega_0) + \Lambda(\Omega_0)} T_b ,
	\label{T_eff_high_T}
\end{equation}
where we have made repeated usage of the fact that
$\Omega_0 \ll T_e, T_b$. It should be
stressed that this result equally applies to all
forms of the bosonic bath, be it Ohmic, sub-Ohmic
or super-Ohmic.

\subsection{Heat current}
The steady-state solution to the probablities $P_n$ can be used in turn to calculate the heat current. Focusing again on the heat current flowing between the bosonic bath and the local phonon, the latter involves the transition rates $W^{\rm b}_{n\to n\pm 1}$ to and from bosonic bath. Explicitly, the heat current takes the form
\begin{equation}
	J_Q = \Omega_0\sum_n P_n \left(W^{\rm b}_{n\to n-1} -
											W^{\rm b}_{n\to n+1}\right),
\end{equation}
where
\begin{eqnarray}
	W^{\rm b}_{n \to n+1} &=& 2\pi(n+1)\Lambda(\Omega_0)n_b(\Omega_0),
	\nonumber \\
	W^{\rm b}_{n \to n-1} &=& 2\pi n
					\Lambda(\Omega_0)\left[1+n_b(\Omega_0)\right].
\end{eqnarray}
Using these rates the expression for the heat current gains the compact form
\begin{equation}
	J_Q = 2\pi \Omega_0 \Lambda(\Omega_0)
	\left[n_{\rm eff}(\Omega_0)-n_b(\Omega_0)
	\right],
	\label{J_Q-master-eq}
\end{equation}
where	$n_{\rm eff}(\Omega_0) = \sum_n n P_n = (e^{\Omega_0/T_{\rm eff}}-1)^{-1}$ is the average occupancy of the localized phonon. Recalling that $T_{\rm eff}$ lies between the temperatures of the bosonic and electronic baths, it is  clear that $J_Q$ is positive (negative) when $T_e > T_b$ ($T_e < T_b$), which gives the correct direction of the heat flow.

The above result provides a transparent picture for the linear dependence of the heat current on the temperature gradient in the high-temperature regime. In this regime, the expression for $J_Q$ can be further approximated using Eq.~(\ref{T_eff_high_T}) for the effective temperature, which results in
\begin{equation}
	J_Q \simeq
				2\pi \frac{\Lambda(\Omega_0) F(\Omega_0)}
				{F(\Omega_0)+\Lambda(\Omega_0)}(T_e-T_b).
\end{equation}
Hence, the linear dependence on the temperature gradient stems from the large occupancy of the localized phonon, which does not saturate with increasing temperature. Segal~\cite{Segal2006}, using the Master equation approach, described a similar linear dependence of the heat flow on the temperature difference. This should be contrasted with a spinless electronic resonant level, that can only be empty or singly occupied.

While Eq.~(\ref{J_Q-master-eq}) properly captures the physics of the high-temperature regime, it fails to produce the required power-law behavior in the low-temperature regime, $T_e,T_b \ll \Omega_0$. Indeed, replacing the Bose-Einstein functions with simple exponents and using Eq.~(\ref{T_eff-via-exponents}) for the effective temperature, the heat current becomes
\begin{equation}
	J_Q \simeq
			2\pi\Omega_0 \frac{\Lambda(\Omega_0) F(\Omega_0)}
				{F(\Omega_0)+\Lambda(\Omega_0)}
		 \left(e^{-\Omega_0/T_e}
			- e^{-\Omega_0/T_b}\right),
\end{equation}
where we have omitted the exponentially small Bose-Einstein distribution functions in the expression for $q$ [see Eq.~(\ref{q-def})]. Thus, the Master equation approach predicts activated low-temperature behavior in place of the correct power-law form.

\subsection{Validity of the Master equation approach}

The fact that the Master equation approach well describes the high-temperature regime but fails at low temperatures is by itself not surprising. Here we can exploit the exact solution to carefully examine the range of validity of the approach and the simple physical picture that it lends. Generally speaking, the quality of the Master equation approach depends on two parameters: (1) the temperatures involved, and (2) the strength of the coupling constants. Naturally, the Master equation approach loses quantitative accuracy as the coupling constants are increased since Fermi's golden-rule corresponds to lowest-order perturbation theory. Nevertheless, as we shall show, the qualitative picture can remain quite accurate even when the coupling constants are increased. To critically examine the Master equation approach we shall focus on the three quantities: (1) The effective temperature $T_{\rm eff}$ defined by the average occupancy of the localized phonon [see Eq.~(\ref{t_eff_from_exact}) below]; (2) the Boltzmann-like distribution of the localized phonon, and (3) the heat current $J_Q$.

\subsubsection{The effective temperature}

As shown above, the Master equation approach predicts an effective temperature $T_{\rm eff}$ which is manifest in the average phonon occupancy $n_{\rm eff}(\Omega_0)$. Similarly, one can define an effective temperature from the exact phononic occupancy $\langle \hat{n}_b \rangle = \langle b^{\dagger}b \rangle$ according to
\begin{equation}
	\langle \hat{n}_b \rangle = \frac{1}{e^{\tilde{\Omega}/T_{\rm eff}}-1},
	\label{t_eff_from_exact}
\end{equation}
where $\tilde{\Omega}$ is the softened phonon frequency, approximated by Eq.~(\ref{omega}). Usage of the softened frequency, rather than the bare one, comes to account for higher-order correction not included in the golden-rule approximation. We emphasis that Eq.~(\ref{t_eff_from_exact}) serves as an ad-hoc definition of an effective temperature, which does not, by itself, imply a Boltzmann-like distribution. We shall examine this latter point in the following subsection.

\begin{figure}[tb]
\centerline{
\includegraphics[width=0.44\textwidth]{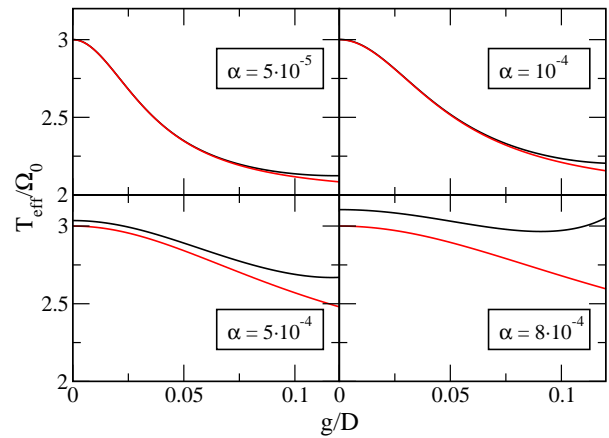}
}\vspace{0pt}
\caption{(Color online) The effective temperature of the localized phonon as a function of the electron-phonon coupling $g$ and for different values of the coupling to the bosonic reservoir $\alpha$. The red line represents the effective temperature calculated based on the Master equation approach while the black line is the effective temperature extracted from the phononic occupancy as given in Eq.~(\ref{t_eff_from_exact}). In this graph $D_{\rm eff}/\Omega_0 = \omega_c/\Omega_0 = 20$, the bosonic bath was taken to be ohmic and the temperatures used are $T_e/\Omega_0=3$, $T_b/\Omega_0=2$, which lie well within the applicatbility temperature range of the Master equation approximation.
}
\label{Fig:t_eff}
\end{figure}

Figure \ref{Fig:t_eff} compares the effective temperature extracted from the exact solution with the Master equation result of Eq.~(\ref{T_eff-via-exponents}), scanning different coupling strengths. There is an excellent agreement at weak coupling which gradually deteriorates as the combined coupling to the two baths is increased. The agreement is controlled by the decay time $\tau$ of Eq.~(\ref{tau}), which depends quadratically on the coupling $g$ to the electronic bath and linearly on the coupling $\alpha$ to the bosonic one. For the strongest coupling considered (right-hand side of the lower-right panel) the deviation between the two curves is of the order of $20\%$, which is still quite moderate.

\subsubsection{The Boltzmann-like distribution}
The Boltzmann-like distribution of the probablities $P_n$ predicts all the moments of the phononic occupancy $\langle n_{\rm eff}^s \rangle = \sum_n n^s P_n$. The ratio between the moments can be expressed using the moments themselves, giving the ratio between the first two moments
\begin{equation}
	\frac{\langle n_{\rm eff}^2 \rangle}{\langle n_{\rm eff} \rangle} =
	2\langle n_{\rm eff} \rangle + 1,
\end{equation}
which does not directly depend on the effective temperature, and is true for every Boltzmann-like distribution, regardless of the value of the temperature. We shall use this ratio as a benchmark for examining how close the phononic distrubtion is to a Boltzmann-like one, as the exact ratio is calculated using the phononic Green functions at steady state
\begin{eqnarray}
	\langle \hat{n}_b \rangle &=& \lim_{t\to\infty}G^{<}_{bb^{\dagger}}(t,t),
	\\
	\langle \hat{n}_b^2 \rangle &=&
	\lim_{t\to\infty}\left\{
	G^{<}_{bb^{\dagger}}(t,t)
	\left[2G^{<}_{bb^{\dagger}}(t,t)+1\right]+
	\left|G^{<}_{bb}(t,t)\right|^2
	\right\}. \nonumber
\end{eqnarray}
The calculated ratio deviates from the Boltzmann-like one due to the presence of the term $G^{<}_{bb}(t,t)$, which is identically zero for thermal distributions.

\begin{figure}[tb]
\centerline{
\includegraphics[width=0.44\textwidth]{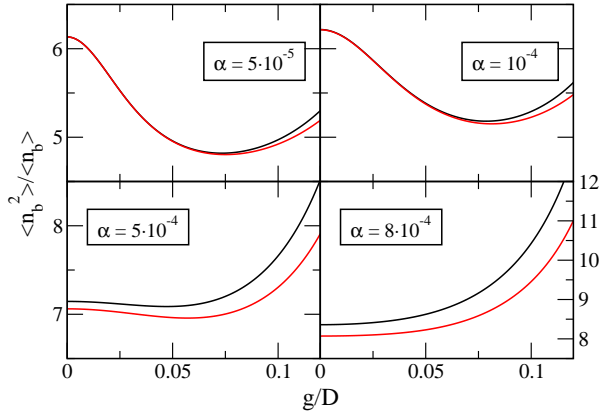}
}\vspace{0pt}
\caption{(Color online)
			The ratio between the second and first moments of the phononic occupation number $\hat{n}_b = b^{\dagger}b$ as a function of the electron-phonon coupling $g$ and for different values of the coupling to the bosonic reservoir $\alpha$. The red line represents the ratio as predicted by the Master equation while the black line is the actual ratio calculated according to the exact solution. In this graph $D_{\rm eff}/\Omega_0 = \omega_c/\Omega_0 = 20$, the bosonic bath was taken to be ohmic and the temperatures used are $T_e/\Omega_0=3$, $T_b/\Omega_0=2$, which lie well within the applicability temperature range of the Master equation approximation.
}
\label{Fig:is_thermal}
\end{figure}

In Fig.~\ref{Fig:is_thermal} we have plotted the exact ratio between the second and first moments of the phononic occupation and compared it to the expected ratio were the distribution was Boltzmann-like. Similarly to the comparison done for the effective temperature, the comparison was done as a function of the coupling strength. For small couplings the exact ratio matches excellently the one expected from a Boltzmann-like distribution, thus confirming the approximation. As the coupling is increased, the exact ratio deviates from the Botlzmann-like prediction, and the non-thermal nature of the distribution becomes prominent. We feel it is important to stress, however, that the distribution is not thermal for any coupling strength.

\subsubsection{The heat current}

\begin{figure}[tb]
\centerline{
\includegraphics[width=0.44\textwidth]{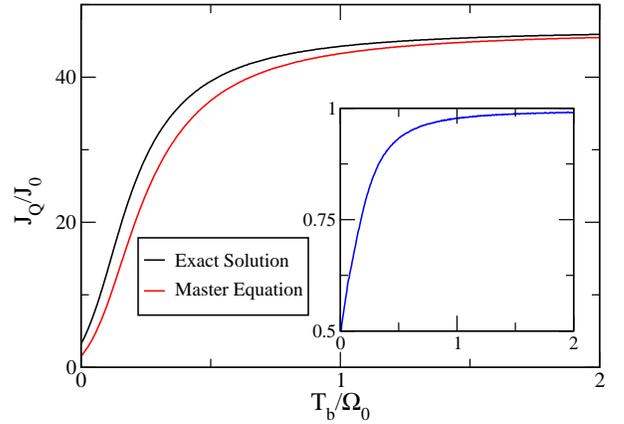}
}\vspace{0pt}
\caption{(Color online)
			A comparison between the exact calculation of
			the heat current between the reservoirs $J_Q$
			and the approximate calculation relying on the
			Master equation approach as given in
			Eq.~(\ref{J_Q-master-eq}). The temperature
			difference between the reservoirs
			is held fixed with $(T_e - T_b)/\Omega_0 = 0.2$
			and both temperatures are changed. Inset: The
			ratio of the approximated result from the exact
			one, showing a larger difference as the temperature
			is lowered. Here $J_Q$ is measured in
			units of the basic heat current
			$J_0 = (4\pi \rho_0 \tilde{g}\Omega_0)^2\alpha$,
			where $2\pi\rho_0 \tilde{g} = 0.1$, and
			$\alpha = 5\! \cdot\! 10^{-4}$ are in the
			weak coupling regime. The cutoffs used
			are $D_{\rm eff}/\Omega_0=\omega_c/\Omega_0 =20$.
}
\label{fig:J_Q_master_eq}
\end{figure}

As a third quantity to which we look in order to examine the validity of the Master equation approximation is the heat current between the two baths $J_Q$. The exact expression for the heat current is given in its integral form in Eq.~(\ref{J_Q-general}) and the expression derived from the Master equation approach given in Eq.~(\ref{J_Q-master-eq}). The comparison of the results of these two expressions is plotted in Fig.~\ref{fig:J_Q_master_eq} where we have scanned different temperatures while holding the temperature difference between the baths constant. In accordance with our previous analysis of the results, the Master equation reproduces quite well the exact result only at high temperatures, and deviating from it significantly as the temperatures are lowered.

\section{Off-resonance condition}
\label{sec:off-resonance}

So far only the resonant case where $\epsilon_d=0$ was considered. Since $\epsilon_d$ can be controlled by using suitable gate voltages, studying its effects on the heat current is of particular interest. In this section we address the case where $\epsilon_d$ is given some nonzero value, breaking the particle-hole symmetry of the model. Staying in accordance with the rest of our discussion, where we have assumed that the bandwidth $D_{\rm eff}$ is the largest energy scale of the system, we are interested in the regime where $|\epsilon_d| \ll D_{\rm eff}$ in addition to $g \ll D_{\rm eff}$. We shall demonstrate that in this regime the leading contribution of $\epsilon_d$ to the heat current is either $\epsilon_d^2 g^2/D_{\rm eff}^4$ or $\epsilon_d^4/D_{\rm eff}^4$.

Before getting to the task of explicitly calculating the heat current for the off-resonant case, we first consider the symmetries of the Hamiltonian of Eq.~(\ref{H}). If $\epsilon_d=0$ then the system holds a particle-hole symmetry, which breaks for nonzero values of $\epsilon_d$.  Writing the Hamiltonian as a function of $\epsilon_d$ and carrying out an particle-hole transformation
\begin{eqnarray}
	c^{\dagger}_k &\to& -c_{-k}, \nonumber \\
	d^{\dagger} &\to& d, \nonumber \\
	b^{\dagger},\;\; \beta^{\dagger}_n &\to& -b^{\dagger},\;\;
	-\beta^{\dagger}_n,
\end{eqnarray}
one finds out that $\mathcal{H}(\epsilon_d) \to \mathcal{H}(-\epsilon_d)$, while the expression for $J_Q$ remains unchanged. Thus we conclude that $J_Q$ is an even function of $\epsilon_d$.

Next we turn to the calculation done in Sec.~\ref{sec:resonance_heat_current} under resonance conditions and aim at adjusting it to the case where $\epsilon_d \neq 0$. Since the $\epsilon_d$ term in the Hamiltonian of Eq.~(\ref{H-bosonic}) is linear in the operators $a_q$ and $a^{\dagger}_q$, one may express its effects on the heat current by appropriate corrections to the bare Green functions pertaining to these operators. We will then be able to use the functions, dressed by the $\epsilon_d$ term, to re-calculate the self-energies  $\Sigma^{r,a}$ of Eq.~(\ref{S^ra}) and $\Sigma^{<,>}$ of Eq.~(\ref{S^<>}).
To this end, we consider a free Hamiltonian to which we add an $\epsilon_d$ term
\begin{equation}
	\mathcal{H}_{\epsilon_d} = \sum_{k>0}\epsilon_k a^{\dagger}_k a_k +
	\tilde{\epsilon}_d\sum_{k>0} \xi_k \left(a^{\dagger}_k + a_k\right),
	\label{H_eps_d}
\end{equation}
where $\tilde{\epsilon}_d=\epsilon_d a$. We will designate by $g^{r,a,<,>}_{0\;k}(t,t')$ the unperturbed Green functions given, in energy space, by
\begin{equation}
	g^{r,a}_{0\;k}(\epsilon) =
       \left[
              \begin{array}{cc}
                    (\epsilon - \epsilon_k \pm i\eta)^{-1}
                    & 0
                    \\ \\
                    0 &
                  - (\epsilon + \epsilon_k \pm i\eta)^{-1}
              \end{array}
       \right],
\label{g0_ra_epsilon_d}
\end{equation}
\begin{equation}
	g^{<,>}_{0\;k}(\epsilon) =
       \pm 2 \pi n_e(\pm \epsilon)
       \left[
              \begin{array}{cc}
                    \delta (\epsilon - \epsilon_k) & 0
                    \\ \\
                    0 & -\delta(\epsilon + \epsilon_k)
              \end{array}
       \right],
\label{g_les_grea_epsilon_d}
\end{equation}
and by $g^{r,a,<,>}_{q,q'}(\omega)$ the dressed functions with respect to the Hamiltonian.

The Hamiltonian can be diagonalized exactly by introducing new bosonic creation and annihilation operators $\tilde{a}_k=a_k+\tilde{\epsilon}_d\xi_k/\epsilon_k$, by which it takes the form
\begin{equation}
	\mathcal{H}_{\epsilon_d} = \sum_{k>0}
						\epsilon_k \tilde{a}^{\dagger}_k \tilde{a}_k -
										\tilde{\epsilon}_d^2\sum_{k>0}\left(
										\frac{\xi_k}{\epsilon_k}\right)^2.
\end{equation}
This is a free Hamiltonian with respect to the $\tilde{a}_k$'s. The correlation functions between the original bosonic operators are related to the correlation functions of the $\tilde{a}_k$'s by the fact that
\begin{equation}
	\langle a^{\dagger}_q(t) a_k (t') \rangle =
	\langle \tilde{a}^{\dagger}_q(t) \tilde{a}_k (t') \rangle +
	\tilde{\epsilon}_d^2\frac{\xi_k\xi_q}{\epsilon_k\epsilon_q}.
	\label{nonresonant_new_bosonic}
\end{equation}
We can thus adjust the lesser and greater Green functions, in the energy domain, by adding the appropriate term proportional to $\delta(\epsilon)$, which reflects the fact that the extra term added by $\epsilon_d$ on the right-hand-side of Eq.~(\ref{nonresonant_new_bosonic}) is independent of time
\begin{eqnarray}
	g^{<,>}_{k,q}(\epsilon) &=& g^{<,>}_{0\; k}(\epsilon)\delta_{k,q} +
	\nonumber \\ &&
	2\pi \tilde{\epsilon}_d^2 \xi_k \xi_q	g^{r}_{0\;k}(0)g^{a}_{0\;q}(0)
	\delta(\epsilon)\left(
	\begin{array}{cc} 1 & 1 \\ 1 & 1
	\end{array} \right).
\end{eqnarray}
We continue to note that the retarded and advanced Green functions remain unchanged by this addition of non-zero $\epsilon_d$, as the extra term is constant and drops out when the commutation relations are taken.

Having arrived at the conclusion that the corrections due to the $\epsilon_d$ term exist only at $\epsilon=0$, we point to the fact that the integrand in the expression for the heat current given in Eq.~(\ref{J_Q-general}) depends directly on $\epsilon$. Therefore, all such terms do not contribute to the heat current, leaving it independent of the value of $\epsilon_d$. As $\epsilon_d$ reflects the energy associated with the charging of the level, one would expect that its value will directly affect the heat current. We understand the independence of the latter on the value of $\epsilon_d$ as a result of the weak-coupling regime $|\epsilon_d|, g \ll D_{\rm eff}$, and expect that as either of these values is increased, $\epsilon_d$ will play a role in determining the heat current. Taking into account that the heat current is an even function of $\epsilon_d$, we conclude our discussion in noting that the leading contribution in this regime is not lower than $(\epsilon_d/D_{\rm eff})^4$ or $(\epsilon_d/D_{\rm eff})^2(g/D_{\rm eff})^2$.

\section{Conclusions}
\label{sec:conclusions}

In this paper we have presented an asymptotically exact calculation of the heat current between a bosonic bath and a fermionic bath, that is mediated by a single molecule. The calculation is based on a mapping of continuous model given by the Hamiltonian of Eq.~(\ref{H}) onto a form quadratic in bosonic operators~\cite{Dora_Halbritter_2009,Dora_2007,Dora_Gulacsi_2008}. This mapping is valid in the weak coupling regime, where $D_{\rm eff}\gg {\rm max}\{g,g^2/\omega_0,|\epsilon_d|\}$, and $D_{\rm eff}$ is the effective electronic bandwidth. This model may describe, under suitable mappings, several physically and experimentally relevant setups, the most relevant being a molecule adsorbed on a surface, a molecular junction and an Aharonov-Bohm interferometer with a molecular device embedded in one of its arms.

The exact calculation yields a Landauer-type expression for the heat current, given in Eq.~(\ref{J_Q-general}). Such an expression stands in accordance with previous works on thermal currents in confined nanostructures.~\cite{Segal2003,Ozpineci2001,Wang2008,Segal2011,Galperin2007,Segal2008,Ojanen2008,Dahr2008}

At low temperatures, the heat current strongly depends on the nature of the bosonic bath. Assuming that the bosonic bath has a power-law form and is characterized by the power $s$, the low-temperature linear-response heat conductance varies as $T^{2+s}$. At high temperatures, however, the heat current depends linearly on the temperature difference between the two baths, regardless of the nature of the power-law governing the bosonic bath. The crossover between the low and high-temperatures regimes is at the scale of the softened vibrational mode frequency $\tilde{\Omega}$ given in Eq.~(\ref{omega}).

The high-temperature behavior, which is markedly different than the transmission through a purely electronic system, is explained by the bosonic nature of vibrational mode, which can be excited to high energies by creating more phonons. This is illustrated by a Master equation analysis of the system, which is perturbative in nature but is justified in the high-temperature and weak coupling regime. In that regime the Master equation approach reproduces the heat current that was calculated exactly previously, and also offers an effective temperature that we assign to the local vibrational mode. It should be stressed that even at that regime it does not have a thermal distribution, and the effective temperature is a useful illustrative approximation.

As our solution is exact only at weak electron-phonon coupling, one would expect different features to appear in the heat current as the interaction strength is increased. It would be interesting to compare our results with such an analysis, and see how our calculation persists into the strong-coupling regime. We leave that to future work.

\begin{acknowledgments}
This work was supported by the US-Israel Binational Science Foundation and by NSF grant DMR 1006684.
\end{acknowledgments}


\begin{thebibliography}{99}
\bibitem{Reviews-of-SMTs}
         For recent reviews see, e.g.,
         {\em Introducing Molecular Electronics},
         edited by G.\ Cuniberti, G.\ Fagas, and
         K.\ Richter,
         Lecture Notes in Physics Vol. 680
         (Springer, New York, 2005);
         M.\ Galperin, M.\ A.\ Ratner, and A.\ Nitzan,
         J.\ Phys.: Condens. Matter {\bf 19},
         103201 (2007).
\bibitem{Wang2008}
			J.-S.\ Wang, J.\ Wang, and J.\ T.\ L\"u,
			Eur.\ Phys.\ J.\ B {\bf 62}, 381-404 (2008).
\bibitem{Phillpot2003}
			D.\ G.\ Cahill, W.\ K.\ Ford, K.\ E.\ Goodson, G.\ D.\ Mahan,
			A.\ Majumdar, H.\ J.\ Maris, R.\ Merlin and S.\ R.\ Phillpot,
			J.\ Appl.\ Phys.\ {\bf 93}, 793 (2003).
\bibitem{Natelson2011}
			D.\ R.\ Ward, D.\ A.\ Corley, J.\ M.\ Tour and D.\ Natelson,
			Nature Nanotechnology {\bf 6}, 33-38 (2011)
\bibitem{Galperin2007}
			M.\ Galperin, A.\ Nitzan, and M.\ A.\ Ratner,
			Phys.\ Rev.\ B {\bf 75}, 155312 (2007).
\bibitem{Entin2010}
		O.\ Entin-Wohlman, Y.\ Imry and A.\ Aharony,
		Phys.\ Rev.\ B {\bf 82}, 115314 (2010).
\bibitem{Entin2012}
		O.\ Entin-Wohlman and A.\ Aharony,
		Phys.\ Rev.\ B {\bf 85}, 085401 (2012).
\bibitem{Ozpineci2001}
			A.\ Ozpineci and S.\ Ciraci,
			Phys.\ Rev.\ B {\bf 63}, 125415 (2001).
\bibitem{Segal2003}
			D.\ Segal, A.\ Nitzan, and P.\ H\"anggi,
			J.\ Chem.\ Phys.\ {\bf 119}, 6840 (2003).
\bibitem{Segal2011}
			L.\ A.\ Wu and D.\ Segal,
			Phys.\ Rev.\ E {\bf 83}, 051114 (2011).
\bibitem{Buttiker2012}
			B.\ Sothmann, R.\ Sanchez, A.\ N.\ Jordan and M.\ B\"uttiker,
			Phys.\ Rev.\ B. {\bf 85}, 205301 (2012).
			B.\ Sothmann and M.\ B\"uttiker,
			EPL {\bf 99}, 27001 (2012).
\bibitem{Glazman-Raikh-88}
         L.\ I.\ Glazman and M.\ Raikh,
         JETP Lett.\ {\bf 47}, 452 (1988).
\bibitem{Franke2012}
			G.\ Schulze, K.\ J.\ Franke, A.\ Gagliardi, G.\ Romano,
			C.\ S.\ Lin, A.\ L.\ Rosa, T.\ A.\ Niehaus, Th.\ Frauenheim,
			A.\ Di Carlo, A.\ Pecchia and J.\ I.\ Pascual,
			Phys.\ Rev.\ Lett.\ {\bf 100}, 136801 (2008);
			G.\ Schulze, K.\ J.\ Franke and J.\ I.\ Pascual,
			New J.\ Phys.\ {\bf 10}, 065005 (2008);
			K.\ J.\ Franke and J.\ I.\ Pascual,
			J.\ Phys.:\ Condens.\ Matter {\bf 24}, 394002 (2012).
\bibitem{Kawai2010}
			M.\ Tsutsui, M.\ Taniguchi, K.\ Yokota and T.\ Kawai,
			Appl.\ Phys.\ Lett.\ {\bf 96}, 103110 (2010).
\bibitem{Segal2008}
			D.\ Segal,
			Phys.\ Rev.\ Lett.\ {\bf 100}, 105901 (2008).
\bibitem{Ojanen2008}
			T.\ Ojanen and A.\ P.\ Jauho,
			Phys.\ Rev.\ Lett.\ {\bf 100}, 155902 (2008).
\bibitem{Ojanen2011}
			T.\ Ruokola and T.\ Ojanen,
			Phys.\ Rev.\ B. {\bf 83}, 045417 (2011).
\bibitem{Dahr2008}
			A.\ Dhar and D.\ Roy,
			J.\ Stat.\ Phys.\ {\bf 125} 801 (2006).
			A.\ Dhar, Adv.\ Phys.\ {\bf 57}, 457 (2008).
\bibitem{Saito2013}
			K.\ Saito and T.\ Kato,
			{\tt arXiv:0706.1234}
\bibitem{comment-on-units}
         We employ units in which $\hbar = k_B = 1$.
\bibitem{Legget1987}
			A.\ J.\ Legget, S.\ Chakravarty, A.\ T.\ Dorsey, M.\ P.\ A.\ Fisher,
			A.\ Garg, and W.\ Zwerger,
			Rev.\ Mod.\ Phys.\ {\bf 59}, 1 (1987).
\bibitem{SvM75}
         D.\ Sherrington and S.\ von Moln\`ar,
         Solid St. Comm. {\bf 16}, 1347 (1975).
\bibitem{cavanagh93}
			R.\ R.\ Cavanagh, E.\ J.\ Heilweil and J.\ C.\ Stephenson,
			Surface Sciente {\bf 299}, 643 (1994).
\bibitem{Gadzuk92}
         J.\ W.\ Gadzuk,
         Phys.\ Rev.\ B {\bf 24}, 1651 (1981);
         E.\ Blaisten-Barojas and J.\ W.\ Gadzuk,
         J.\ Chem.\ Phys.\ {\bf 97}, 862 (1992).
\bibitem{VSA2012}
         Y.\ Vinkler, A.\ Schiller, and N.\ Andrei,
			Phys.\ Rev.\ B {\bf 85}, 035411 (2012).
\bibitem{Lozanne93-5}
		H.\ L.\ Edwards, Q.\ Niu and A.\ L.\ de Lozanne,
		Appl.\ Phys.\ Lett.\ {\bf 63}, 1815 (1993);
		H.\ L.\ Edwards, Q.\ Niu, G.\ A.\ Georgakis
		and A.\ L.\ deLozanne,
		Phys.\ Rev.\ B {\bf 52}, 5714 (1995).
\bibitem{Ritchie2009}
		J.\ R.\ Prance, C.\ G.\ Smith, J.\ P.\ Griffiths,
		S.\ J.\ Chorley, D.\ Anderson, G.\ A.\ C.\ Jones,
		I.\ Farrer and D.\ A.\ Ritchie,
		Phys.\ Rev.\ Lett.\ {\bf 102}, 146602 (2009).
\bibitem{LF62}
         I.\ G.\ Lang and Yu.\ A.\ Firsov,
         Zh.\ Eksp.\ Teor.\ Fiz.\ {\bf 43}, 1843 (1962)
         [Sov.\ Phys.\ JETP {\bf 16}, 1301 (1963)].
\bibitem{Dora_Halbritter_2009}
         B.\ D\'{o}ra and A.\ Halbritter,
         Phys.\ Rev.\ B. {\bf 80}, 155402 (2009).
\bibitem{Dora_2007}
         B.\ D\'{o}ra,
         Phys.\ Rev.\ B. {\bf 75}, 245113 (2007).
\bibitem{Dora_Gulacsi_2008}
         B.\ D\'{o}ra and M.\ Gul\'acsi,
         Phys.\ Rev.\ B. {\bf 78}, 165111 (2008).
\bibitem{Schoeller2001}
		W.\ Hofstetter, J.\ K\"onig and H.\ Schoeller,
		Phys.\ Rev.\ Lett.\ {\bf 87}, 156803 (2001).
\bibitem{Keldysh1965}
			L.\ V.\ Keldysh,
			Sov.\ Phys.\ JETP {\bf 20}, 1018 (1965).
\bibitem{Haldane81}
         F.\ D.\ M.\ Haldane,
         J.\ Phys. C {\bf 14}, 2585 (1981).
\bibitem{Langreth76}
			D.\ C.\ Langreth, in
			{\em Linear and Nonlinear Electron Transport in Solids}
			(Plenum Pres, New York, 1976), vol.\ 17 of {\em Nato Advanced
			Study Institute, Series B: Physics}, eds.\ J.\ T.\	Devreese and
			V.\ E.\ van Doren.
\bibitem{AS-E1}
         See, e.g.,
         {\em Handbook of Mathematical Functions},
         eds. M.\ Abramowitz and I.\ A.\ Stegun
         (Dover, New York, 1972), Chapter 5.
\bibitem{AS-zeta}
         See, e.g.,
         {\em Handbook of Mathematical Functions},
         eds. M.\ Abramowitz and I.\ A.\ Stegun
         (Dover, New York, 1972), Chapter 23.
\bibitem{Hanggi2011}
			F.\ Zhan, S.\ Denisov and P.\ H\"anggi,
			Phys.\ Rev.\ B {\bf 84}, 195117 (2011).
\bibitem{AS-psi}
         See, e.g.,
         {\em Handbook of Mathematical Functions},
         eds. M.\ Abramowitz and I.\ A.\ Stegun
         (Dover, New York, 1972), Chapter 6.
\bibitem{Segal2006}
			D.\ Segal,
			Phys.\ Rev.\ B {\bf 73}, 205415 (2006).
\bibitem{Flesnberg2010}
			M.\ Leijnse, M.\ R.\ Wegewijs and K.\ Flensberg,
			Phys.\ Rev. B {\bf 82}, 045412 (2010).
\end{thebibliography}
\end{document}